# On the Optimality of a Class of LP-based Algorithms [*]

Amit Kumar[†]     Rajsekar Manokaran[‡]     Madhur Tulsiani[§]     Nisheeth K. Vishnoi[¶]

**Abstract**

In this paper we will be concerned with a class of packing and covering problems which includes Vertex Cover and Independent Set. Typically, one can write an LP relaxation and then round the solution. For instance, for Vertex Cover one can obtain a 2-approximation via this approach. On the other hand, Khot and Regev [KR08] proved that, assuming the Unique Games Conjecture (UGC), it is NP-hard to approximate Vertex Cover to within a factor better than $2 - \varepsilon$ for any constant $\varepsilon > 0$. From their, and subsequent proofs of this result, it was not clear why this LP relaxation should be optimal.

The situation was akin to Maximum Cut, where a natural SDP relaxation for it was proved by Khot *et al.* [KKMO07] to be optimal assuming the UGC. A beautiful result of Raghavendra [Rag08] explains why the SDP is optimal (assuming the UGC). Moreover, his result generalizes to a large class of constraint satisfaction problems (CSPs). Unfortunately, we do not know how to extend his framework so that it applies for problems such as Vertex Cover where the constraints are *strict*.

In this paper, we explain why the simple LP-based rounding algorithm for the Vertex Cover problem is optimal assuming the UGC. Complementing Raghavendra's result, our result generalizes to a class of *strict*, covering/packing type CSPs. We first write down a natural LP relaxation for this class of problems and present a simple rounding algorithm for it. The key ingredient, then, is a dictatorship test, which is parametrized by a *rounding-gap* example for this LP, whose completeness and soundness are the LP-value and the rounded value respectively.

To the best of our knowledge, ours is the first result which proves the optimality of LP-based rounding algorithms systematically.

[*]This work was done while the authors were at Microsoft Research India, Bangalore

[†]Dept. of Computer Science and Engineering, IIT Delhi, New Delhi, email: `amitk@cse.iitd.ac.in`

[‡]Department of Computer Science, Princeton University, email: `rajsekar@cs.princeton.edu`. Supported by NSF grants 0830673, 0832797, 528414

[§]Institute for Advanced Study, Princeton, email: `madhurt@cs.berkeley.edu`. Supported by NSF grant 0832797 and IAS sub-contract no. 00001583.

[¶]Microsoft Research India, Bangalore, email: `nisheeth.vishnoi@gmail.com`

# 1 Introduction

In this paper we will be interested in a class of packing and covering problems. The class of covering problems we study (packing problems can be defined similarly) contains the set of problems characterized by the following integer program: given non-negative $A \in \mathbb{R}^{m \times n}, b \in \mathbb{R}^m, c \in \mathbb{R}^n$, the goal is to find a vector $x \in \{0, \ldots, q-1\}^n$ such that $Ax \geq b$ which minimizes $\sum_i c_i x_i$. In addition, each row of $A$ has at most $k$ non-zero entries and $k$ and $q$ are assumed to be constants. We will show that, assuming the Unique Games Conjecture, a natural linear programming relaxation (not the naive one) for such a problem, along with a simple rounding scheme, is the best one can hope for algorithmically. We start with a classic covering problem that falls in this class: the minimum vertex cover problem (Vertex Cover). The equivalent packing problem is the maximum independent set problem (Independent Set).

**Vertex Cover.** In the Vertex Cover problem, one is given a graph $G(V, E)$ with non-negative weights $\{w_v\}_{v \in V}$ for the vertices and the goal is to find a subset of minimum weight such that every edge has at least one end point in this set. We may assume that $\sum_{v \in V} w_v = 1$. This problem is NP-hard and there is a very simple 2-approximation algorithm known for it. One way to obtain this approximation, which will be relevant to our work, is to solve the following natural linear programming (LP) relaxation.

$$\mathsf{lp}_{\mathsf{vc}}(G) \stackrel{\text{def}}{=} \quad \text{minimize} \quad \sum_{v \in V} w_v x_v$$
$$\text{subject to} \quad \forall_{e=uv \in E} \quad x_u + x_v \geq 1$$
$$\forall_{v \in V} \quad x_v \geq 0$$

Figure 1: LP for Vertex Cover

Given an optimal solution to this LP, the rounding scheme selects vertices whose variables have value at least $1/2$. This makes sure that the picked set is a vertex cover and the cost is no more than twice $\mathsf{lp}_{\mathsf{vc}}(G)$.

On the inapproximability side, Dinur and Safra proved that Vertex Cover is NP-hard to approximate to within a factor of 1.36. Khot and Regev [KR08] gave some evidence that factor 2 might be the best possible approximation for Vertex Cover by showing that if there is an algorithm for Vertex Cover which achieves a factor better than 2, then the Unique Games Conjecture (UGC) is false. However, it is not clear from their hardness reduction why this simple minded LP is the best one.

One can think of Vertex Cover as a *strict* version of a constraint satisfaction problem (CSP). While in a CSP, the objective is to satisfy as many constraints as possible, in Vertex Cover, the objective is to find a solution satisfying *all* constraints while minimizing a different objective. Although the difference seems superficial, the algorithmic implications for the two class of problems are quite different. Let us take the example of Maximum Cut, the simplest and one of the most interesting CSP. Given a graph $G(V, E)$, the Maximum Cut problem asks for a partition $(S, \bar{S})$ of the vertices maximizing the number of edges going across the cut. For the Maximum Cut problem, the natural LP relaxation is known to give a strictly weaker approximation than a simple semidefinite programming (SDP) relaxation [GW95]. Further, the factor obtained by the SDP is optimal assuming the UGC as was shown in Khot *et al.* [KKMO07]. A beautiful result of Raghavendra [Rag08] manages to generalize the Maximum Cut result of [KKMO07] to a very general class of constraint satisfaction problems. In particular, he considers a generic way to write a SDP relaxation, and shows that the approximation ratio achived by the SDP is optimal assuming the UGC.

However, Raghavendra's result does not seem to apply for Vertex Cover even though it is possible to think of Vertex Cover as a CSP in the following way: consider an instance $G = (V, E)$ of Vertex Cover. We have boolean variables $x_v$ for each $v \in V$. For each vertex $v$, we have a constraint $\Phi_v(x_v)$ which is 1 if $x_v$ is 1 (which corresponds to vertex $v$ being picked in the vertex cover), and 0 otherwise. For each edge $e = uv$, we have a constraint $\Psi_e(x_u, x_v)$ which is 0 if at least one of $x_u$ or $x_v$ is 1, and 1 otherwise. The goal is to minimize the weighted sum $\sum_v w_v \Phi_v(x_v) + 2 \cdot \sum_{e=uv \in E}(w_u + w_v) \cdot \Psi_e(x_u, x_v)$. It is easy to argue that any optimal solution $x^\star$ to this CSP must satisfy $\Psi_e(x_u^\star, x_v^\star) = 0$ for all edges $uv \in E$. Hence, any optimal solution to this CSP also yields an optimal vertex cover solution. Given this reduction from VERTEX COVER to a weighted CSP, one can imagine using the techniques of Raghavendra [Rag08] to come up with a similar hardness result for VERTEX COVER. However, such an argument fails to work for the following reason: the CSP corresponding to VERTEX COVER has a very special relation between the edge constraints and the vertex constraints – the weight of each edge constraint is more than the sum of the weights of the corresponding vertex constraints. The reduction from UGC to such a CSP (which has vertex and edge constraints as defined above) will construct hard instances of such CSPs which may not preserve the required relationship between the weights of vertex and edge constraints. This problem seems to remain whenever we have CSPs where some constraints are *strict*, i.e., they cannot be violated. This leaves the understanding of problems with strict constraints as the VERTEX COVER problem somewhat unsettled.

## 1.1 Our Work

In this paper, assuming UGC, we present an optimal LP-based algorithm for a class of packing and covering problems. First, we describe the class of covering problems we will study. This description contains the problems arising from the integer programming formulation described before. Here we describe the $q = 2$ case. [1]

**Strict Monotone Constraint Satisfaction Problem.** A problem $\Pi$ is said to be a $k$-strict and monotone CSP ($k$-sm-CSP) if it consists of a set of vertices $V$ with non-negative weights $\{w_v\}_{v \in V}$ on them, a set of hyperedges of size at most $k$ and for every hyperedge $e \in E$, a constraint $A_e$. These constraints have the following two properties natural for $\{0, 1\}$-covering problems.

1. The constraints are *strict*: the constraint corresponding to every hyperedge has to be satisfied.

2. The constraints are *monotone*: given a feasible solution (a subset of vertices), adding more vertices to the solution keeps it feasible.

The objective is to find a boolean assignment to the vertices so as to satisfy all the hyperedge constraints and minimize the weight of vertices set to 1 (selected). We denote by $\Pi_{A_1,...,A_t}$ a problem in the class $k$-sm-CSP if, for every instance of $\Pi_{A_1,...,A_t}$ and every hyperedge $e$ of this instance, $A_e$ is one of the constraints $\{A_1, \ldots, A_t\}$. It is easy to observe that VERTEX COVER is a special 2-sm-CSP.

**An LP for a $k$-sm-CSP problem.** One can define the following LP relaxation for any problem in the class $k$-sm-CSP. As written, it may not be immediately clear that it is an LP. We will prove this in Section 2.1 and for the benefit of the reader, we also show in the appendix (Section 6.1) how the standard VERTEX COVER LP can be equivalently written in this form. This relaxation is inspired by the Sherali-Adams [SA90] relaxation and plays a crucial role in our result.

Here, for a hyperedge $e = \{v_1, \ldots, v_l\}$, `ConvexHull`($A_e$) denotes the convex hull of all assignments $\sigma \in \{0, 1\}^l$ which satisfy the constraint $A_e$. For an instance $\mathcal{I}$, let $\mathsf{lp}(\mathcal{I})$ denote the optimum of the LP of Figure 2 for $\mathcal{I}$. Also, let $\mathsf{opt}(\mathcal{I})$ denote the value of the optimal integral solution for $\mathcal{I}$.

---

[1]We note that for the $\{0, 1\}$ case, HYPERGRAPH VERTEX COVER are all the problems one gets in this class. Though the $q$-ary case, for $q > 2$, is more general, all important ideas in this paper are equally well captured by the binary case. Given this, for improved readability of this paper, the presentation for most of this paper is chosen to be for the $\{0, 1\}$ case. In Section 5, we outline our results in the general setting of the $q$-ary alphabets.

$$\text{lp}(\mathcal{I}) \stackrel{\text{def}}{=} \quad \text{minimize} \quad \sum_{v \in V} w_v x_v \tag{1}$$

$$\text{subject to} \quad \forall_{e=(v_1,v_2,\ldots,v_l) \in E} \quad (x_{v_1}, x_{v_2}, \ldots, x_{v_l}) \in \texttt{ConvexHull}(A_e) \tag{2}$$

$$\forall_{v \in V} \quad 0 \leq x_v \leq 1 \tag{3}$$

Figure 2: LP for $k$-sm-CSP

**Main Result.** The main contribution of this paper is to integrate a class of covering and packing problems in the same conceptual framework as done by Raghavendra for CSPs [Rag08]. We give a rounding algorithm called ROUND (see Figure 3) for the LP of Figure 2. For an instance $\mathcal{I}$ of $\Pi$ we first compute $x^\star$, the optimal LP-solution for $\mathcal{I}$. For a parameter $\varepsilon > 0$, which should be ignored for this discussion, let $\text{round}(\mathcal{I}, x^\star, \varepsilon)$ denote the value of ROUND. We then show how to start with an instance $\mathcal{I}$ of $\Pi$ and give a UNIQUE GAMES-based reduction for $\Pi$ whose soundness and completeness are roughly $\text{lp}(\mathcal{I})$ and $\text{round}(\mathcal{I}, x^\star, \varepsilon)$ respectively. We also show that ROUND (unconditonally) achieves an approximation ratio equal to the integrality gap, up to an arbitrarily small additive constant, of the LP relaxation. This can be seen as an analogue of the result of Raghavendra and Steurer [RS09]. We assume throughout that $k = O(1)$ as our reductions are polynomial-sized when $k$ is a constant.

**Theorem 1.1.** *Let $\Pi$ be a $k$-sm-CSP, $\varepsilon > 0$ for $k = O(1)$, and $\mathcal{I}$ be an instance of $\Pi$. Then for every $\delta > 0$, assuming the UGC, it is NP-hard to distinguish instances of $\Pi$ with optimal less than $\text{lp}(\mathcal{I}) + \varepsilon + \delta$ from those with optimal more than $\text{round}(\mathcal{I}, x^\star, \varepsilon) - \delta$. Here $x^\star$ is the optimal LP solution for $\mathcal{I}$.*

As a corollary we can deduce the following: we set $\delta \stackrel{\text{def}}{=} \varepsilon$ and note that for any instance $\mathcal{I}$, $\text{round}(\mathcal{I}, x^\star, \varepsilon)$ is at most $\text{opt}(\mathcal{I})$.

**Corollary 1.2.** *Let $\Pi$ be in the class $k$-sm-CSP and $\mathcal{I}$ be an instance of $\Pi$. Then for every $\delta > 0$, assuming the UGC, it is NP-hard to distinguish instances of $\Pi$ with optimal less than $\text{lp}(\mathcal{I}) + 2\delta$ from those with optimal more than $\text{opt}(\mathcal{I}) - \delta$.*

The class $k$-sm-CSP falls under the class of *covering problems*. One can also define a similar class of problems for *packing problems*, namely Packing-$k$-sm-CSP. The only differences would be (1) we would be interested in maximizing the total weight of vertices set to 1, and (2) the constraints would be *downward* monotone, i.e., removing some vertices from the solution would still keep it feasible. All our results translate to the setting of packing problems in the natural manner and we omit the details.

### 1.2 Overview and Techniques

In this section we outline the proof of Theorem 1.1. We explain the main ideas with the canonical example of VERTEX COVER. Unlike previous work, we will be interested in instances where ROUND fails i.e., where the value $\text{round}(\mathcal{I}, x^\star, \varepsilon)$ is much larger than $\text{lp}(\mathcal{I})$. Note that this is a weaker condition than the instance being an integrality gap instance (as considered previously for CSPs), where the value $\text{opt}(\mathcal{I})$ is much larger than $\text{lp}(\mathcal{I})$.

In particular, for VERTEX COVER, ROUND performs poorly on a graph with a single edge, which is not an integrality gap instance because the integer optimum and the LP optimum are both equal to 1. One optimal LP solution is to set a value of $1/2$ for both the end points, and thus, satisfying the constraint corresponding to this edge. ROUND would then end up choosing both the end points in the cover, thus, being off by a factor 2 in the objective. We show a reduction from UNIQUE GAMES that uses this instance to obtain a factor $2 - \varepsilon$ hardness for VERTEX COVER. The reduction produces the Khot-Regev graph although our analysis is conceptually simpler and uses the, by now standard, Invariance Principle. Note that the inapproximability obtained is at least as good as the integrality gap of the LP because the optimal solution is at least as good as the one output by ROUND.

Given an instance of VERTEX COVER along with an LP solution, we prove that the best rounding scheme, assuming the UGC, is one which sets the variables to 1 or 0 by solely looking at the value of the variables. We do so by picking a UNIQUE GAMES instance and replacing each vertex of the unique game with several blocks of vertices, one for every distinct value the LP solution takes (one can think of having one block for every variable in the LP, but we will later need to identify variables taking the same value). We then introduce constraints across the blocks inspired by the LP solution such that if the UNIQUE GAMES instance has a good labeling, then there is a way to choose a fraction of each of the blocks such that all the edges are covered. Further, the fraction selected in each block has measure exactly equal to the total value of the corresponding variables in the LP solution.

Next, we show that if the UNIQUE GAMES instance has no good solution, then any solution can be converted to a solution that either completely chooses a block or ignores it completely while losing only a small additive error in the objective. We use the Invariance Principle and gaussian stability estimates to prove that if a particular block is partially selected, then, there are a bunch of constraints that are completely contained in the unselected part. Thus, given that the solution satisfies every constraint (covers all edges), the partially selected blocks might as well have been completely unselected without violating any constraint. If we think of selecting a block as setting the corresponding variable to 1 and not selecting as setting it to 0, this naturally corresponds to a rounding of the LP solution solely based on the value of each variable: exactly the space of solutions ROUND searches over.

As it is, this does not say much since the LP value could assign completely distinct values to all the variables. However, we can always convert such an LP solution to one that takes only a constant ($1/\varepsilon$) number of distinct values while losing a small additive error ($\varepsilon$) in the objective. This naturally gives a rounding scheme, which as the first step *discretizes* the LP solution to have a small number of distinct values before trying out every possible rounding of the values. The rounding scheme remains exactly the same for the general version: solve the linear program, fix the solution to one with a small number of distinct values and then exhaustively search over all possible roundings of the values.

## 1.3 Further Discussion and Related Work

**LP inspired hardness results.** There are several problems for which the best known inapproximability results have been obtained as follows: first construct integrality gap instances for the standard LP relaxations for these problems and then use these instances as guides for constructing hardness reductions based on standard complexity theoretic assumptions. These reductions yield inapproximability ratios quite close to the actual integrality gaps. Examples include ASYMMETRIC $k$-CENTER [CGH[+]04], GROUP STEINER TREE [HK03] and AVERAGE FLOW-TIME ON PARALLEL MACHINES [GK07]. Assuming UGC, our result proves hardness of a large class of problems in a similar spirit. However, instead of explicitly constructing integrality gap examples for such problems, we give a more direct and intuitive proof that the integrality gap is close to the actual hardness of such problems. We note that the only other result for LPs similar in flavor as ours, though unrelated, is that of [MNRS08] for MULTI-WAY CUT.

**Unique Games Conjecture.** Since its inception, the UGC of Khot [Kho02] has been used to obtain a host of inapproximability results [Kho02, KV05, CKK[+]06, KKMO07, KR08, MNRS08, GMR08, Rag08, CGM09] and, it implies optimal hardness of approximation results for problems such as MAXIMUM CUT [KKMO07] and VERTEX COVER [KR08, BK09].

Very recently, for the problem of VERTEX COVER, a similar analysis using the Invariance Principle was also proved by Austrin, Khot and Safra [AKS09]. They were motivated by the problem of proving hardness of approximating VERTEX COVER on bounded degree graphs where as the goal of this paper is to establish the optimality of LP-based algorithms for problems similar to VERTEX COVER.

## 1.4 Rest of the paper.

In Section 2, we give a detailed outline of the proof of Theorem 1.1. In Section 2.1 we formally define the class of problems we shall be interested in. In Section 2.2, we outline the properties of the dictatorship test which leads to the proof of Theorem 1.1. We describe the rounding algorithm and prove its optimality in Section 3. We give all the details of the construction and the properties of the dictatorship test in Section 4. This requires us to prove some properties of a feasible solution to the LP, which we describe in Section 4.1. The actual reduction from UGC to a problem in the class $k$-sm-CSP builds on the construction of the dictatorship function in a standard manner – we describe this in the appendix (Section 6.5). We describe the setting of the $q > 2$ case in Section 5.

## 2 Proof of Theorem 1.1

In this section we will reduce the task of proving Theorem 1.1 to constructing a dictatorship test. Once we have stated the claim for the dictatorship test, it would be standard to convert it into a UGC -based hardness result and deduce Theorem 1.1. The soundness of this dictatorship test, in turn, relies on the rounding algorithm ROUND of Figure 3 which is described in Section 3. We start with some necessary preliminaries to describe the problems and the LP we will consider for them.

### 2.1 Preliminiaries

Given distinct $x, y \in \{0, 1\}^k$, we say $y \geq x$, if for every $i \in [k]$, $y_i \geq x_i$. Further, a subset $A \subseteq \{0, 1\}^k$ is said to be *upward monotone* if for every $x \in A$, and every $y$ such that $y \geq x$, it follows that $y \in A$.

**Definition 2.1** (The class $k$-sm-CSP). Let $k$ be a positive integer. An instance of type $k$-sm-CSP is given by

$$\mathcal{I} = (V, E, \{A_e\}_{\{e \in E\}}, \{w_v\}_{v \in V}) \text{ where :}$$

- $V = \{v_1, v_2, \ldots, v_n\}$ denotes a set of variables/vertices taking values over $\{0, 1\}$ along with non-negative weights such that $\sum_{v \in V} w_v = 1$.
- $E$ denotes a collection of hyperedges, each on at most $k$ vertices. For each hyperedge $e \in E$, there is a constraint $A_e$ which is an upward monotone set denoting the set of accepted configurations of the vertices in $e$.

The objective is to find an assignment $\Lambda : V \mapsto \{0, 1\}$ for the vertices in $V$ that minimizes $\sum_{v \in V} w_v \Lambda(v)$ such that for each $e = (v_1, v_2, \ldots, v_l)$, $(\Lambda(v_1), \ldots, \Lambda(v_l)) \in A_e$. A $k$-sm-CSP $\Pi$ is given by upward monotone sets $\{A_1, \ldots, A_t\}$. Every instance of $\Pi$ is supposed to have its constraints for the hyperedges to be one of $\{A_1, \ldots, A_t\}$. For simplicity, we will assume that the size of each hyperedge is exactly $k$. Our proofs also hold when hyperedges have size at most $k$.

**Definition 2.2.** For a set $A \subseteq \{0, 1\}^k$, define ConvexHull($A$) as the convex hull of elements in $A$. Note that any element $x \in$ ConvexHull($A$) can be expressed as $\sum_{\sigma \in A} \lambda_\sigma \cdot \sigma$, where $\lambda_\sigma$ can be thought of as giving a probability distribution over elements of $A$.

The LP relaxation for $k$-sm-CSP appears in Figure 2. On an instance $\mathcal{I}$ and a feasible solution $x$ to LP($\mathcal{I}$), we will let val($\mathcal{I}, x$) to denote the objective of LP($\mathcal{I}$). We now show that this indeed is an LP by explicitly writing the constraints (2) as linear constraints. Consider an edge $e = (u_1, \ldots, u_k) \in E$. We define variables $\lambda_\sigma^e$, where

$\sigma \in \{0, 1\}^k$ and it varies over all elements of $A_e$. The constraint (2) can now be written as

$$(x_{u_1}, x_{u_2}, \ldots, x_{u_k}) = \sum_{\sigma \in A_e} \lambda^e_\sigma \sigma$$

$$\forall_{\sigma \in A_e} \; \lambda^e_\sigma \geq 0$$

$$\sum_{\sigma \in A_e} \lambda^e_\sigma = 1.$$

In the appendix (Section 6.1), we show that for the VERTEX COVER and HYPERGRAPH VERTEX COVER problems, this LP is at least as strong as the standard LP relaxations for these problems.

## 2.2 Dictatorship Function

We now describe the properties of the dictatorship test. In fact, we prefer to refer to it as a dictatorship *function* as it takes some inputs and has an output rather than an acceptance predicate.

Our dictatorship function $\text{DICT}^\Pi_{\mathcal{I}, x, m}(r, \delta)$ takes as input an instance $\mathcal{I}$ of a problem $\Pi$, a solution $x$ to the LP relaxation (Figure 2) for this instance, and parameters $m, r$ and $\delta$. $r$ would be the label set size of the UNIQUE GAMES instance we combine this dictatorship function with to get the actual reduction for Theorem 1.1 and can be ignored for now. $m$ is an upper bound on the number of distinct values the entries in the vector $x$ are allowed to take. We will often refer to $\varepsilon$ as $1/m-1$. $\delta$ is a *smoothening* parameter, essential for the application of the Invariance Principle, and can be ignored for now. The dictatorship function outputs another instance $\mathcal{D}$ of $\Pi$. The vertex set $V(\mathcal{D})$ of this instance would be $[m] \times \{0, 1\}^r$ and we defer the precise description of the output instance for Section 4.2. Recall that $\text{round}(\mathcal{I}, x, \varepsilon)$ denotes the value of the rounding algorithm $\text{ROUND}(\mathcal{I}, x, \varepsilon)$. Even though we will describe the algorithm in the next section, for now, it is sufficient to know that the algorithm produces an integral solution for $\mathcal{I}$ from a feasible solution $x$ to $\text{LP}(\mathcal{I})$ after it has perturbed $x$ and made sure that its variables take at most $m$ distinct values. In particular, $\text{round}(\mathcal{I}, x, \varepsilon) \leq \text{opt}(\mathcal{I})$. Now, we can state the main claim regarding our dictatorship function.

**Lemma 2.3.** *(Informal Version) The dictatorship function $\text{DICT}^\Pi_{\mathcal{I}, x, m}(r, \delta)$ has the following properties:*

1. Completeness: *There are $r$ "dictator" assignments to $V(\mathcal{D})$ each of which satisfy all the constraints of $\mathcal{D}$ and each costs at most $\text{val}(\mathcal{I}, x) + \delta$.*

2. Soundness: *Every assignment which is "far from being a dictator" assignment and satisfies all the constraints in $\mathcal{D}$ must have cost at least $\text{round}(\mathcal{I}, x, \varepsilon) - \delta$.*

We leave the precise definition of "dictator" and "far from being a dictator" for Section 4.2. The proof of this lemma also appears in Section 4.2. For a fixed instance $\mathcal{I}$, to maximize the gap between $\text{val}(\mathcal{I}, x)$ and $\text{round}(\mathcal{I}, x, \varepsilon)$ one should use $x = x^\star$, the optimal solution to $\text{LP}(\mathcal{I})$. The problem is that, then, we are not guaranteed that the variables in $x^\star$ takes at most $m$ distinct values. But this is easy: since the variables take value between 0 and 1, one can just *bucket* them into bins of width $\varepsilon$ and lose only an additional $\varepsilon$ in the completeness in the lemma above. Hence, we can obtain the following corollary.

**Corollary 2.4.** *The dictatorship function $\text{DICT}^\Pi_{\mathcal{I}, x^\star, m}(r, \delta)$ has the following properties:*

1. Completeness: *There are $r$ dictator assignments to $V(\mathcal{D})$ each of which satisfy all the constraints of $\mathcal{D}$ and each costs at most $\text{lp}(\mathcal{I}) + \varepsilon + \delta$.*

2. Soundness: *Every assignment which is "far from being a dictator" assignment and satisfies all the constraints in $\mathcal{D}$ must have cost at least $\text{round}(\mathcal{I}, x^\star, \varepsilon) - \delta$.*

*Here $x^\star$ is the optimal LP solution for $\mathcal{I}$.*

The corollary above can be converted to a Unique Games-based hardness result for $\Pi$. The following is just a reformulation of Theorem 1.1 and will be proved formally in Section 6.5.

**Theorem 2.5.** *Let $\Pi$ be a $k$-sm-CSP, $\varepsilon > 0$ for $k = O(1)$, and $\mathcal{I}$ be an instance of $\Pi$. Then for every $\delta > 0$, given an input instance $\mathcal{J}$ of $\Pi$, assuming the UGC, it is NP-hard to distinguish between the following:*

1. Completeness: $\mathsf{opt}(\mathcal{J}) \leq \mathsf{lp}(\mathcal{I}) + \varepsilon + \delta$.

2. Soundness: $\mathsf{opt}(\mathcal{J}) \geq \mathsf{round}(\mathcal{I}, x^\star, \varepsilon) - \delta$.

*Here $x^\star$ is the optimal LP solution for $\mathcal{I}$.*

## 3 The Rounding Algorithm and its Optimality

In this section we describe our rounding algorithm ROUND and prove that it achieves the integrality gap unconditionally. Theorem 1.1 implies that it is not possible to beat this integrality gap assuming the UGC. Hence, this algorithm is optimal.

**The algorithm.** The algorithm will use a parameter $\varepsilon$. We assume without loss of generality that $1/\varepsilon$ is an integer. We first define a way of perturbing a solution $x$ to $\mathsf{LP}(\mathcal{I})$ (Figure 2) such that the number of distinct values the variables of $x$ take is at most $1/\varepsilon + 1$.

**Definition 3.1.** *Given an $x$ such that $0 \leq x_u \leq 1$ for all $u \in V$, and a parameter $\varepsilon > 0$, define $x^\varepsilon$ as follows – for each $u \in V$, let $k_u$ be the integer satisfying $k_u \varepsilon < x_u \leq (k_u + 1)\varepsilon$, then $x_u^\varepsilon \stackrel{\text{def}}{=} (k_u + 1)\varepsilon$ (if $x_u = 0$, we define $x_u^\varepsilon$ to be 0 as well).*

In other words, $x^\varepsilon$ is obtained from $x$ by rounding up each coordinate to the nearest integral multiple of $\varepsilon$ (note that this value will not exceed 1 because $1/\varepsilon$ is an integer). First we observe the following simple fact.

**Fact 3.2.** *Let $x$ be a feasible solution to $\mathsf{LP}(\mathcal{I})$. Then*

1. *$x^\varepsilon$ is feasible for $\mathsf{LP}(\mathcal{I})$.*

2. *$\mathsf{val}(\mathcal{I}, x^\varepsilon) \leq \mathsf{val}(\mathcal{I}, x) + \varepsilon$.*

The proof this fact appears in the appendix (Section 6.2).
   The algorithm ROUND is described in Figure 3. This algorithm takes as input an instance $\mathcal{I}$, a feasible solution $x$ to $\mathsf{LP}(\mathcal{I})$ and a parameter $\varepsilon > 0$. We denote $\mathsf{round}(\mathcal{I}, x, \varepsilon)$ as the value of the integral solution returned by $\mathsf{ROUND}(\mathcal{I}, x, \varepsilon)$. First, the algorithm perturbs $x$ to $x^\varepsilon$ to make sure that the number of distinct values taken by the variables in $x^\varepsilon$ is at most $m = O(1/\varepsilon)$, which is to be thought of as a (large) constant. Thus, the variables fall into $m$ buckets and now, the rounding algorithm goes over all possible assignments to these constantly many buckets and outputs the assignment with the least cost.

**The optimality of the rounding algorithm.** We quickly observe that ROUND achieves the integrality gap.

**Lemma 3.3.** *Let $\gamma^*(\Pi)$ be the worst-case approximation ratio (integrality gap) achieved by the LP relaxation for the problem $\Pi$, i.e., $\gamma^*(\Pi) \stackrel{\text{def}}{=} \sup_{\mathcal{I}}(\mathsf{opt}(\mathcal{I})/\mathsf{lp}(\mathcal{I}))$, where the supremum is taken over all instances $\mathcal{I}$ of $\Pi$. Then, for any given instance $\mathcal{J}$, optimal LP solution $x$ and $\varepsilon > 0$, $\mathsf{round}(\mathcal{J}, x, \varepsilon) \leq \gamma^*(\Pi) \cdot (\mathsf{opt}(\mathcal{J}) + \varepsilon)$.*

*Proof.* Consider an input $(\mathcal{J}, x, \varepsilon)$ to the algorithm ROUND. We define a new instance $\mathcal{J}'$ of $\Pi$ as follows: the set of variables $V' = \{0, \ldots, 1/\varepsilon + 1\}$ and hyperedge set $E' = \{(i_1, \ldots, i_k) \mid (v_1, \ldots, v_k) \in E \text{ and } x^\varepsilon_{u_j} = i_j \cdot \varepsilon \text{ for all } j \in [k]\}$. We take the weight $w_i$ of $i \in V'$ to be $\sum_{x^\varepsilon_v = i\varepsilon} w_v$ and take constraint $A_{e'}$ for an edge $e' \in E'$ to be the same as $A_e$ for the corresponding edge in $e \in E$. Then, since $\text{ROUND}(\mathcal{J}, x, \varepsilon)$ searches over all feasible assignments to variables in $\mathcal{J}'$, we get that $\text{round}(\mathcal{J}, x, \varepsilon) = \text{opt}(\mathcal{J}')$. Hence, we get

$$\text{round}(\mathcal{J}, x, \varepsilon) = \text{opt}(\mathcal{J}') \leq \gamma^*(\Pi) \cdot \text{lp}(\mathcal{J}') \overset{\text{Fact 3.2–(2)}}{\leq} \gamma^*(\Pi) \cdot (\text{lp}(\mathcal{J}) + \varepsilon) \leq \gamma^*(\Pi) \cdot (\text{opt}(\mathcal{J}) + \varepsilon).$$

□

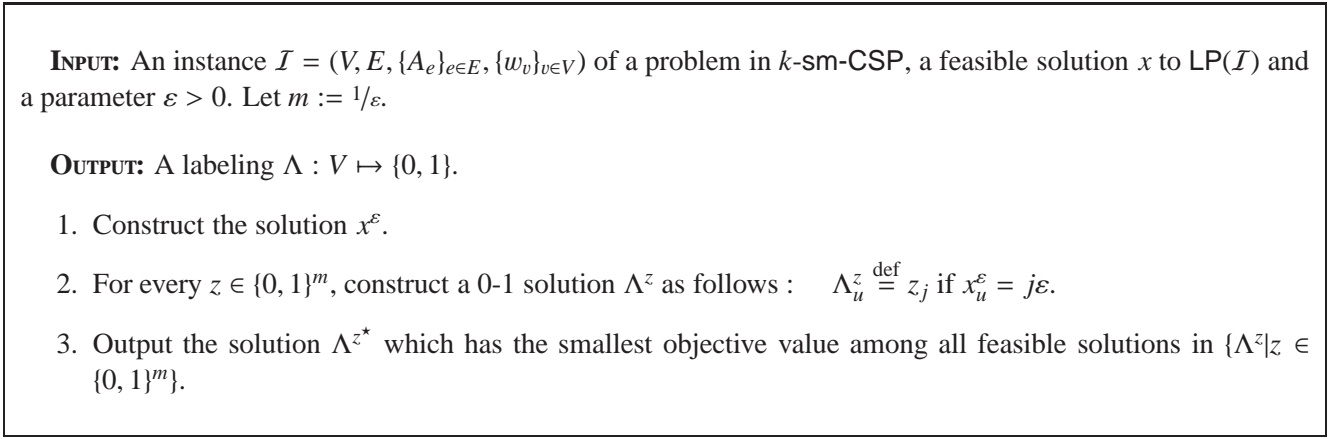

**INPUT:** An instance $\mathcal{I} = (V, E, \{A_e\}_{e \in E}, \{w_v\}_{v \in V})$ of a problem in $k$-sm-CSP, a feasible solution $x$ to LP($\mathcal{I}$) and a parameter $\varepsilon > 0$. Let $m := 1/\varepsilon$.

**OUTPUT:** A labeling $\Lambda : V \mapsto \{0, 1\}$.

1. Construct the solution $x^\varepsilon$.

2. For every $z \in \{0, 1\}^m$, construct a 0-1 solution $\Lambda^z$ as follows: $\Lambda^z_u \overset{\text{def}}{=} z_j$ if $x^\varepsilon_u = j\varepsilon$.

3. Output the solution $\Lambda^{z^\star}$ which has the smallest objective value among all feasible solutions in $\{\Lambda^z | z \in \{0, 1\}^m\}$.

Figure 3: Algorithm ROUND

**Deducing the Khot-Regev result.** As an application of Theorem 1.1, we prove the result of Khot and Regev [KR08] which states that assuming UGC, it is NP-hard to get $(k-\varepsilon)$-approximation algorithm for the $k$-HYPERGRAPH VERTEX COVER problem for any constant $\varepsilon > 0$. Consider the following instance $\mathcal{I}$ of $k$-HYPERGRAPH VERTEX COVER – we are given a set $V$ of size $k$, and there is only one hyperedge in $E$, namely, the set of all vertices in $V$. The weight of every vertex is $1/k$. Consider the solution $x$ which assigns value $1/k$ to all variables $x_u$. It is easy to check that it is feasible to our LP relaxation. The value of the solution $x$ is $1/k$. . Let us now see how the algorithm ROUND($\mathcal{I}, x, \varepsilon$), where $\varepsilon < 1/k$, rounds the solution $x$. All entries in $x^\varepsilon$ will still be same. Hence, the rounding algorithm will consider only two options – either pick all vertices in $V$, or do not pick any vertex. Since the latter case yields an infeasible solution, it will output the set $V$, which has value 1. Theorem 1.1 now implies that assuming UGC, it is NP-hard to distinguish between instances of $k$-HYPERGRAPH VERTEX COVER where the optimal value is at most $1/k - 2\varepsilon$ from those where the optimal value is more than $1 - \varepsilon$. Note that the integrality gap of LP($\mathcal{I}$) is 1. Still we are able to argue hardness of $k$-HYPERGRAPH VERTEX COVER problem starting from such an instance because the algorithm ROUND performs poorly on this instance. In this sense, the statement of Theorem 1.1 is stronger than that of Corollary 1.2.

## 4 Dictatorship Function

In this section we give details of the dictatorship function. The dictatorship function takes as input an instance $\mathcal{I}$ of the problem, a feasible solution $x$ to lp($\mathcal{I}$) and parameters $m, \delta$ and $r$. The entries in the solution $x$ take at most $m$ distinct values. The parameter $r$ is related to the label size in the UNIQUE GAMES instance that is used for the reduction, and can be assumed to be an arbitrary quantity for this discussion. The output of the dictatorship function is another instance $\mathcal{D}$ of the same problem. The vertex set of this instance is $[m] \times \{0, 1\}^r$, which we can think of as *m blocks*. The hyperedges in $\mathcal{D}$ are constructed roughly as follows – for every hyperedge $e$ in the input instance $\mathcal{I}$, we pick $r$ hyperedges from the set $A_e$ by a randomized process. We use these $r$ hyperedges to construct one hyperedge in $\mathcal{D}$ which spans across $k$ suitably chosen blocks in $\mathcal{D}$. The set of hyperedges in $\mathcal{D}$ consist of hyperedges which can be obtained in this manner with strictly positive probability. Note that even though these probabilities are not explicitly required for the description of $\mathcal{D}$, they are still useful for the soundness analysis of the dictatorship function.

Given a hyperedge $e$ in the instance $\mathcal{I}$, the probabilities with which we select $r$ edges from $A_e$ are obtained by using the fact that $x$ is a feasible solution to $\mathsf{lp}(\mathcal{I})$. This allows us to express $x$ as a convex combination of elements in $A_e$ and the coefficients in this convex combination give us the requisite probabilities. The analysis of the dictatorship function uses the Invariance Principle, which requires that the minimum non-zero probability of an element in $A_e$ is not too small, and a suitably defined "correlation graph" is connected. We achieve these two properties by using the parameter $\delta$ to "smoothen" the probabilties obtained from the solution $x$. We now define these probability distributions and their smoothenings more formally below.

## 4.1 Probability Distribution from the LP solution

We define the notion of smoothening the distribution obtained by restricted $x$ to a hyperedge $e$.

**Definition 4.1.** Let $x$ be a feasible solution to $\mathsf{LP}(\mathcal{I})$. For a hyperedge $e$, let $x|_e$ be the vector obtained by restricting $x$ to only those vertices which belong to $e$. $P_e^x$ shall denote the probability distribution over the vectors in $\{0,1\}^k$ corresponding to elements of $A_e$ that arises from the fact that $x|_e \in \texttt{ConvexHull}(A_e)$. For a parameter $\delta > 0$, define a random function $R^\delta : \{0,1\}^k \mapsto \{0,1\}^k$ which satisfies the following property independently for each $i$:

$$R^\delta(x)_i \stackrel{\text{def}}{=} \begin{cases} x_i & \text{with probability } 1 - \delta \\ 1 & \text{with probability } \delta \end{cases}$$

Now define a distribution $M_\delta(P_e^x)$ over $\{0,1\}^k$ where

$$M_\delta(P_e^x)(z) \stackrel{\text{def}}{=} \mathop{\mathbf{P}}_{z', R^\delta}\left[R^\delta(z') = z\right],$$

where $z'$ is drawn according to distribution $P_e^x$ and $R^\delta$ from the space of function from $\{0,1\}^k$ to $\{0,1\}^k$ as described above.

**Remark 4.2.** Note that for a hyperedge $e$, the support of $P_e^x$ is a subset of $A_e$. Since $A_e$ is upward monotone, the same fact holds for $M_\delta(P_e^x)$ as well.

**Definition 4.3.** Let $P$ be a distribution on $\{0,1\}^k$. Define $\text{margin}_i(P), 1 \leq i \leq k$ as the following distribution on $\{0,1\}$:

$$\text{margin}_i(P)(b) \stackrel{\text{def}}{=} \sum_{b_1,\ldots,b_{i-1},b_{i+1},\ldots,b_k} P(b_1,\ldots,b_{i-1},b,b_{i+1},\ldots,b_k).$$

The following fact, whose proof is easy to check, shows that the marginals of the distributions $P_e^x$ and $M_\delta(P_e^x)$ on a single coordinate can be easily expressed in terms of $x$.

**Fact 4.4.** *Let $e = (u_1, \ldots, u_k)$ by a hyperedge in an instance $\mathcal{I}$. Then, $\mathbf{E}_{\text{margin}_i(P_e^x)}(b) = x_{u_i}$, and $\mathbf{E}_{\text{margin}_i(M_\delta(P_e^x))}(b) = (1-\delta)x_{u_i} + \delta$.*

**Fact 4.5.** *For a hyperedge $e$, let $\alpha$ denote $\min_\omega\{P_e^x(\omega) : P_e^x(\omega) > 0\}$. Then $\min_\omega\{M_\delta(P_e^x(\omega)) : M_\delta(P_e^x(\omega)) > 0\}$ is at least $\delta^k \cdot \alpha$.*

A feasible solution $x$ satisfies the property that $(x_{u_1}, \ldots, x_{u_k})$ can be expressed a convex combination of vectors in $A_e$ for every $e = (u_1, \ldots, u_k) \in E$. The following lemma shows that under some conditions on $x$, the non-zero coefficients of this convex combination are large enough. This shall be useful in our application of the Invariance Principle. We defer the proof to appendix (Section 6.3).

**Lemma 4.6.** *Let $A$ be a upward monotone subset of $\{0,1\}^k$ and let $(x_1, x_2, \ldots, x_k) \in$ ConvexHull$(A)$ such that each $x_u$ is an integral multiple of $\varepsilon$ (assume that $1/\varepsilon$ is an integer). Then, there is a distribution $P$ over $A$ such that the minimum probability of any atom in $P$ is at least $\frac{\varepsilon}{2^k!}$.*

## 4.2 Description and Properties of the Dictatorship Function

In this section, we first describe the parameters of the dictatorship function and the output instance that it produces. Then we give a more formal statement of Lemma 2.3 in Lemma 4.9 and Theorem 4.10. We then prove these and hence, complete the proof of Lemma 2.3 and Corollary 2.4. For $0 \le \ell \le 1$, let $z \leftarrow \mu_\ell$ denote a string drawn from the $\ell$-biased distribution on $\{0,1\}^r$.

**Parameters of** $\text{Dict}^\Pi_{I,x,m}(r, \delta)$.

1. **Starting Instance.** An instance $I = (V = \{u_1, \ldots, u_n\}, E, \{A_e\}_{e \in E}, \{w_u\}_{u \in V})$ of $\Pi$.

2. **Solution to LP$(I)$.** A vector $x = (x_{u_1}, \ldots, x_{u_n})$ which is a feasible solution to LP$(I)$.

3. **Multiplicity Parameter.** $m$ denotes the number of distinct values in the vector $x$ and let $p_1, \ldots, p_m$ denote these distinct values. Let $B: V \mapsto \{1, 2, \ldots, m\}$ such that $B(u) := b$ whenever $x_u = p_b$ for some $b \in \{1, \ldots, m\}$.

4. **Smoothening Parameter.** $\delta > 0$. For a feasible solution $x$ to LP$(I)$ and for every edge $e \in E$, we use this parameter to define the distribution $M_\delta(P_e^x)$. The solution $x$ shall satisfy the conditions of Lemma 4.6, and, hence, Fact 4.5 implies that the minimum probability of an atom in $M_\delta(P_e^x)$ is at least $\frac{\delta^k \varepsilon}{2^k!}$.

5. **Repetition Parameter.** A positive integer $r$. ($r$ would be the size of the label set of the Unique Games instance to be used in the hardness reduction.)

**The Output Instance.** $\text{Dict}^\Pi_{I,x,m}(r, \delta)$ outputs the following instance $\mathcal{D}$ of $\Pi$

1. **Vertex Set.** The vertex set of $\mathcal{D}$ is $V(\mathcal{D}) := [m] \times \{0,1\}^r$.

2. **Weights on Vertices.** Let $\tilde{p}_b \stackrel{\text{def}}{=} p_b(1 - \delta) + \delta$ for $b \in \{1, 2, \ldots, m\}$. The weight of a vertex $(b, y) \in \mathcal{D}$,

$$w_\mathcal{D}((b, y)) \stackrel{\text{def}}{=} \mu_{\tilde{p}_b}(y) \sum_{u \in V | x_u = p_b} w_u.$$

Note that since, $\sum_{u \in V} w_u = 1$, $\sum_{v \in V(\mathcal{D})} w_\mathcal{D}(v) = 1$.

3. **Hyper-edges and Constraints.** The set of constraints in $\mathcal{D}$ is the union of every constraint output with positive probability by the following procedure:

   (a) Pick a random hyperedge $e = (u_1, u_2, \ldots, u_k)$ from $I$ and let $A_e$ denote the constraint corresponding to it.
   
   (b) Sample $r$ independent copies $(z^j_{u_1}, z^j_{u_2}, \ldots, z^j_{u_k})^r_{j=1}$ from the distribution $M_\delta(P_e^x)$.

(c) Let the corresponding hyperedge in $\mathcal{D}$ be

$$\left(\left(B(u_1), (z^1_{u_1}, z^2_{u_1}, \ldots, z^r_{u_1})\right), \left(B(u_2), (z^1_{u_2}, z^2_{u_2}, \ldots, z^r_{u_2})\right), \ldots, \left(B(u_k), (z^1_{u_k}, z^2_{u_k}, \ldots, z^r_{u_k})\right)\right)$$

and the constraint corresponding to it be $A_e$.

We now define the notions "dictator" and "far from being dictator", and then give formal statements of Lemma 2.3.

**Definition 4.7.** A set $S \subseteq [m] \times \{0, 1\}^r$ is said to be a dictator if there exists an $i \in [r]$ such that $S = \{(b, z) : z_i = 1\}$.

For a $S \subseteq [m] \times \{0, 1\}^r$, define $S_b \stackrel{\text{def}}{=} S \cap (b, \{0, 1\}^r)$. Further, define functions $f^S_b : \{0, 1\}^r \mapsto \{0, 1\}$ to be the indicator function of the *complement* of $S_b$. We refer to a subset of vertices of $\mathcal{D}$ of the form $(b, \{0, 1\}^r)$ as the *hypercube corresponding to $b$*. Given $i \in [r], b \in [m]$, the degree $d$-influence of the $i$-th coordinate of $f^S_b$, $\text{Inf}^{\leq d}_i(f^S_b)$ with respect to the measure $\mu_{\tilde{p}_b}$ is defined in Section 6.4.1.

**Definition 4.8.** Given $\tau, d \geq 0$, a set $S \subseteq [m] \times \{0, 1\}^r$ is said to be $(\tau, d)$-*pseudo-random* if for every $b \in [m]$ and every $i \in [r]$, $\text{Inf}^{\leq d}_i(f^S_b) \leq \tau$.

Now we state the completeness and soundness properties of the dictatorship function.

**Lemma 4.9.** *(Completeness.) Suppose $S \subseteq [m] \times \{0, 1\}^r$ is a dictator. Then $S$ satisfies all the constraints of $\mathcal{D}$ and $w_{\mathcal{D}}(S) \leq \text{val}(\mathcal{I}, x) + \delta$.*

Observe that if we take the optimal solution $x$ of $\text{lp}(\mathcal{I})$ and use the corresponding perturbed solution $x^\varepsilon$ as input to the dictatorship function, then Lemma 4.9 combined with Fact 3.2 implies the first statement in Corollary 2.4.

**Theorem 4.10.** *(Soundness.) For every small enough $\delta > 0$, there exists a $d, \tau$ such that if $S \subseteq [m] \times \{0, 1\}^r$ satisfies all the constraints of $\mathcal{D}$ and is $(\tau, d)$-pseudo-random, then*

$$w_{\mathcal{D}}(S) \geq \text{round}(\mathcal{I}, x, \varepsilon) - \delta.$$

The construction of the dictatorship function can be extended to give a reduction from UNIQUE GAMES to a problem in the class $k$-sm-CSP. The completeness and soundness of the dictatorship function translate to that for the reduction, which allows us to prove Theorem 2.5. We describe this reduction in the appendix (Section 6.5).

**Proofs of Completeness and Soundness Properties.** Now we give proofs of Lemma 4.9 and Theorem 4.10.

**Proof of Lemma 4.9.** Let $i$ be such that $S = \{(b, z) : z_i = 1\}$. For any hyperedge $e \in E$, recall that the support of $M_\delta(P^x_e)$ is a subset of $A_e$. Hence, any $(z_{u_1}, z_{u_2}, \ldots, z_{u_k}) \leftarrow M_\delta(P^x_e)$ always satisfies the constraint $A_e$. Thus, the set $S$ satisfies every constraint of the instance $\mathcal{D}$. The total weight of the set $S$ is

$$w_{\mathcal{D}}(S) = \sum_{b \in [m]} \sum_{z|z_i=1} \mu_{\tilde{p}_b}(z) \sum_{u|x_u=p_b} w_u = \sum_{b \in [m]} \tilde{p}_b \sum_{u|x_u=p_b} w_u = \sum_u w_u x_u (1 - \delta) + \delta \leq \text{val}(\mathcal{I}, x) + \delta.$$

□

We now prove the soundness of the dictatorship test. The crux of the proof is to show that given a subset $S \subseteq [m] \times \{0, 1\}^r$ satisfying all the constraints in $\mathcal{D}$, one can find a subset $J \subseteq [m]$ such that $T_J \stackrel{\text{def}}{=} \cup_{b \in J}(b, \{0, 1\}^r)$ satisfies all the constraints in $\mathcal{D}$ and $w_{\mathcal{D}}(T_J) \leq w_{\mathcal{D}}(S) + \delta$.

**Proof of Theorem 4.10.** Define

$$J \stackrel{\text{def}}{=} \left\{b \mid \mathop{\mathbf{E}}_{z \leftarrow \mu_{\tilde{p}_b}} f^S_b(z) \leq \delta\right\}.$$

Fix a particular hyperedge $e = (u_1, u_2, \ldots, u_k) \in E$ of the instance $\mathcal{I}$. Let

$$R_e \stackrel{\text{def}}{=} \{u_i \in e \,|\, B(u_i) \notin J\}.$$

We will deduce the following fact from the Invariance Principle of Mossel [Mos08]. We give a detailed description of the Invariance Principle in the appendix (Section 6.4.1). The proof of the following fact is also deferred to the appendix (Section 6.4).

**Fact 4.11.**
$$\mathbf{E}\left[\Pi_{u \in R_e} f^S_{B(u)}(z^{(u)})\right] > 0.$$

In the expectation above, the argument $z^{(u)}$ of $f^S_{B(u)}$ is chosen with the $\tilde{p}_{B(u)}$-biased measure on $\{0,1\}^r$ independently for each $u \in R_e$. Since $f^S_b \in \{0,1\}$ and, in particular non-negative, for each $b \in [m]$, this fact implies that there is a hyperedge in $\mathcal{D}$ that queries the hypercubes corresponding to $(u_1, u_2, \ldots, u_k)$ at $z^{(u_1)}, \ldots, z^{(u_k)}$ respectively such that for every $u \in R_e$, $f^S_{B(u)}(z^{(u)}) = 1$. Since $f^S_b$ is the indicator of the complement of $S_b$, this means that the constraint was satisfied solely by the hypercubes in $J$. Thus, all the constraints of $\mathcal{D}$ would also be satisfied by the set $T_J$. Note that since $T_J$ selects hypercubes that are already $1 - \delta$ fraction covered by $S$, $w_\mathcal{D}(T_J) \leq w_\mathcal{D}(S) + \delta$.

Further, since $T_J$ either completely chooses a particular hypercube or completely ignores it, this corresponds to one of the assignments that the rounding algorithm iterates through on the input $(\mathcal{I}, x, \varepsilon)$. This implies that $\mathsf{round}(\mathcal{I}, x, \varepsilon)$ is at most $w_\mathcal{D}(T_J)$. This completes the proof of the theorem. □

## 5 Extension to larger alphabets

In this section, we show how our results can be easily extended to the case when variables take values from a larger alphabet $[q] = \{0, \ldots, q-1\}$.

**Problem Definition.** Given $x, y \in [q]^k$, we say that $y \geq x$, if, $y_i \geq x_i$ for all $i$, $1 \leq i \leq k$. A set $A \subseteq [q]^k$ is said to be *upward monotone* if for every $x \in A$, and every $y$ such that $y \geq x$, it follows that $y \in A$. For sake of brevity, we assume that the alphabet size, $q$, is implicit in the definition below.

**Definition 5.1** (The class $k$-sm-CSP). Let $k$ be a positive integer. An instance of type $k$-sm-CSP is given by

$$\mathcal{I} = (V, E, \{A_e\}_{\{e \in E\}}, \{w_v\}_{v \in V}) \text{ where :}$$

- $V = \{v_1, v_2, \ldots, v_n\}$ denotes a set of variables/vertices taking values over $[q]$ along with non-negative weights such that $\sum_{v \in V} w_v = 1$.

- $E$ denotes a collection of hyperedges, each on at most $k$ vertices. For each hyperedge $e \in E$, there is a constraint $A_e$ which is an upward monotone set denoting the set of accepted configurations of the vertices in $e$.

The objective is to find an assignment $\Lambda : V \mapsto [q]$ for the vertices in $V$ that minimizes $\sum_{v \in V} w_v \Lambda(v)$ such that for each $e = (v_1, v_2, \ldots, v_l)$, $(\Lambda(v_1), \ldots, \Lambda(v_l)) \in A_e$. A $k$-sm-CSP $\Pi$ is given by upward monotone sets $\{A_1, \ldots, A_t\}$. Every instance of $\Pi$ is supposed to have its constraints for the hyperedges to be one of $\{A_1, \ldots, A_t\}$. As in the binary case, we will assume that the size of each hyperedge is exactly $k$. Our proofs continue to hold when hyperedges have size at most $k$.

**LP relaxation**  We now give an LP relaxation for a problem in $k$-sm-CSP. The following definition allows us to map values in $[q]$ to vectors whose coordinates lie between 0 and 1.

**Definition 5.2.** Let $\Delta_q$ denote the set of vectors $\{(z_0, \ldots, z_{q-1}) : z_i \geq 0$ for all $i \in [q]$ and $\sum_{i \in [q]} z_i = 1\}$. There is a natural mapping $\Psi_q : \Delta_q \mapsto [q]$ defined as $\Psi_q((z_0, \ldots, z_{q-1})) = \sum_{i \in [q]} z_i \cdot i$. Let $e_i$, for $i \in [q]$, be the unit vector in $\mathbb{R}^q$ which has value 1 at coordinate $i$, and 0 elsewhere. It is easy to check that $\Delta_q$ is the convex hull of the vectors $\{e_i : i \in [q]\}$. It follows that a vector $x \in \Delta_q$ can also be thought of as a probability distribution over $[q]$.

**Definition 5.3.** Given an integer $i \in [q]$, define $\Phi_q(i)$ as the vector $e_i \in \mathbb{R}^q$. Given a sequence $\sigma \in [q]^k$, for some parameter $k$, define $\Phi_q(\sigma) = (\Phi_q(\sigma_1), \ldots, \Phi_q(\sigma_k))$. Note that $\Phi_q(\sigma)$ is a vector in $\mathbb{R}^{q \cdot k}$.

The LP relaxation for an instance $\mathcal{I}$ of a problem $\Pi \in k$-sm-CSP is described in Figure 4.

$$\mathsf{lp}(\mathcal{I}) \stackrel{\text{def}}{=} \quad \text{minimize} \quad \sum_{v \in V} w_v \Psi_q(x_v) \tag{4}$$

$$\text{subject to} \quad \forall_{e=(v_1, v_2, \ldots, v_k) \in E} \quad (x_{v_1}, x_{v_2}, \ldots, x_{v_k}) \in \mathtt{ConvexHull}(A_e) \tag{5}$$

$$\forall_{v \in V} \quad x_v \in \Delta_q \tag{6}$$

Figure 4: LP for $k$-sm-CSP

Here, $\mathtt{ConvexHull}(A_e)$ is the convex hull of the set $\{\Phi_q(\sigma) : \sigma$ is a satisfying assignment for $A_e\}$. It is easy to check that this indeed is a linear program. Given a solution $x$ to $\mathsf{LP}(\mathcal{I})$, let $\mathsf{val}(\mathcal{I}, x)$ denote the objective function value for $x$.

**The Rounding Algorithm.**  We now describe the rounding algorithm. The algorithm uses a perturbation parameter $\varepsilon$. We first argue that we can perturb a feasible solution to $\mathsf{LP}(\mathcal{I})$ such that the number of distinct (vector) values taken by the variables are small. This perturbation will not affect the objective value significantly. We shall assume without loss of generality that $1/\varepsilon$ is an integer.

**Definition 5.4.** For a parameter $\varepsilon > 0$, define $\Delta_q^{\varepsilon,i}$, $0 \leq i < q$, as the set of points $z \in \Delta_q$ satisfying the following conditions – (1) $z_0, \ldots, z_{i-1}$ are multiples of $\varepsilon$, and (2) $z_{i+1} = \cdots = z_{q-1} = 0$. Observe that $z_i$ must equal $1 - \sum_{j=0}^{i-1} z_j$. Let $\Delta_q^\varepsilon$ denote $\bigcup_i \Delta_q^{\varepsilon,i}$.

It is easy to check that $|\Delta_q^\varepsilon|$ is at most $1/(\varepsilon+1)^q$. We now show how a vector $x \in \Delta_q$ can be perturbed to a vector in $\Delta_q^\varepsilon$.

**Definition 5.5.** Let $a \in [0, 1]$ be a real number. Define $a^\varepsilon$ as the smallest multiple of $\varepsilon$ greater than or equal to $a$. Consider $x \in \Delta_q$. Let $i$ be the largest integer such that $x_0^\varepsilon + \cdots + x_{i-1}^\varepsilon \leq 1$. Then, define $x^\varepsilon$ to be the vector $(x_0^\varepsilon, \ldots, x_{i-1}^\varepsilon, 1 - \sum_{j=0}^{i-1} x_j^\varepsilon, 0, \ldots, 0) \in \Delta_q^{\varepsilon,i}$.

The rounding algorithm is described in Figure 5. Let $\mathsf{round}_q(\mathcal{I}, x, \varepsilon)$ denote the objective value of the solution returned by $\mathsf{ROUND}_q(\mathcal{I}, x, \varepsilon)$.
Since $m_q$ is $O(1/\varepsilon^q)$, the running time of $\mathsf{ROUND}_q$ is $O(\mathsf{poly}(n^k, 1/\varepsilon^q))$. We state the following fact without proof.

**Fact 5.6.** *Let $x$ be a feasible solution to $\mathsf{LP}(\mathcal{I})$. Then*

1. *$x^\varepsilon$ is feasible for $\mathsf{LP}(\mathcal{I})$.*

2. *$\mathsf{val}(\mathcal{I}, x^\varepsilon) \leq \mathsf{val}(\mathcal{I}, x) + \varepsilon \cdot q^2$.*

> **Input:** An instance $\mathcal{I} = (V, E, \{A_e\}_{e \in E}, \{w_v\}_{v \in V})$ of a problem in $k$-sm-CSP, a feasible solution $x$ to $\mathsf{LP}(\mathcal{I})$ and a parameter $\varepsilon > 0$. Let $m_q$ denote $|\Delta_q^\varepsilon|$.
>
> **Output:** A labeling $\Lambda : V \mapsto [q]$.
>
> 1. Construct the solution $x^\varepsilon$.
>
> 2. Let $I$ denote the set $\Delta_q^\varepsilon$ arranged in some order.
>
> 3. For every $z \in [q]^{m_q}$, construct an integral solution $\Lambda^z$ as follows: $\quad \Lambda_u^z \stackrel{\text{def}}{=} z_j$ if $x_u^\varepsilon = I_j$.
>
> 4. Output the solution $\Lambda^{z^*}$ which has the smallest objective value among all feasible solutions in $\{\Lambda^z | z \in [q]^{m_q}\}$.

Figure 5: Algorithm $\mathsf{ROUND}_q$

**Probability Distribution from a Feasible Solution.**

**Definition 5.7.** Let $x$ be a feasible solution to $\mathsf{LP}(\mathcal{I})$. For a hyperedge $e$, let $x|_e$ be the vector obtained by restricting $x$ to only those vertices which belong to $e$. $P_e^x$ shall denote the probability distribution over vectors in $\Delta_q^k$ corresponding to elements of $A_e$ that arises from the fact that $x|_e \in \mathtt{ConvexHull}(A_e)$. For a parameter $\delta > 0$, define a random function $R^\delta : \Delta_q^k \mapsto \Delta_q^k$ which satisfies the following property independently for each $i$:

$$R^\delta(x)_i \stackrel{\text{def}}{=} \begin{cases} x_i & \text{with probability } 1 - \delta \\ e_{q-1} & \text{with probability } \delta \end{cases}$$

Now, define a distribution $M_\delta(P_e^x)$ over $\Delta_q^k$ where

$$M_\delta(P_e^x)(z) \stackrel{\text{def}}{=} \mathbf{P}_{z', R^\delta}\left[R^\delta(z') = z\right],$$

where $z'$ is drawn according to distribution $P_e^x$ and $R^\delta$ from the space of function from $\Delta_q^k$ to $\Delta_q^k$ as described above.

**Definition 5.8.** Given a vector $p \in \Delta_q$, let $\mu_p$ be the corresponding distriution over $[q]$ (as described in Definition 5.2).

**The Dictatorship Function.** We now describe the parameters of the dictatorship function and the output instance that it produces.
**Parameters of** $\mathrm{DICT}_{\mathcal{I}, x, m}^\Pi(r, \delta)$.

1. Starting Instance. An instance $\mathcal{I} = (V = \{u_1, \ldots, u_n\}, E, \{A_e\}_{e \in E}, \{w_u\}_{u \in V})$ of $\Pi$.

2. Solution to $\mathsf{LP}(\mathcal{I})$. $x = (x_{u_1}, \ldots, x_{u_n})$ which is a feasible solution to $\mathsf{LP}(\mathcal{I})$.

3. Multiplicity Parameter. $m$ denotes the number of distinct (vector) values in the vector $x$ and let $p_1, \ldots, p_m$ denote these distinct values. Note that $p_i \in \Delta_q$ for all $i$. Let $B : V \mapsto \{1, 2, \ldots, m\}$ such that $B(u) := b$ whenever $x_u = p_b$ for some $b \in \{1, \ldots, m\}$.

4. Smoothening Parameter. $\delta > 0$. For a feasible solution $x$ to $\mathsf{LP}(\mathcal{I})$ and for every edge $e \in E$, we use this parameter to define the distribution $M_\delta(P_e^x)$.

5. **Repetition Parameter.** A positive integer $r$. ($r$ would be the size of the label set of the UNIQUE GAMES instance to be used in the hardness reduction.)

**The Output Instance.**
$\text{DICT}_{\mathcal{I},x,m}^{\Pi}(r,\delta)$ outputs the following instance $\mathcal{D}$ of $\Pi$

1. **Vertex Set.** The vertex set of $\mathcal{D}$ is $V(\mathcal{D}) := [m] \times [q]^r$.

2. **Weights on Vertices.** Let $\tilde{p}_b \stackrel{\text{def}}{=} (1-\delta) \cdot p_b + \delta \cdot e_{q-1}$ for $b \in \{1, 2, \ldots, m\}$. The weight of a vertex $(b, y) \in \mathcal{D}$,

$$w_{\mathcal{D}}((b,y)) \stackrel{\text{def}}{=} \mu_{\tilde{p}_b}(y) \sum_{u \in V | x_u = p_b} w_u.$$

Note that since, $\sum_{u \in V} w_u = 1$, $\sum_{v \in V(\mathcal{D})} w_{\mathcal{D}}(v) = 1$.

3. **Hyper-edges and Constraints.** The set of constraints in $\mathcal{D}$ is the union of every constraint output with positive probability by the following procedure:

    (a) Pick a random hyperedge $e = (u_1, u_2, \ldots, u_k)$ from $\mathcal{I}$ and let $A_e$ denote the constraint corresponding to it.
    
    (b) Sample $r$ independent copies $(z_{u_1}^j, z_{u_2}^j, \ldots, z_{u_k}^j)_{j=1}^r$ from the distribution $M_\delta(P_e^x)$.
    
    (c) For each $j = 1, \ldots, r$, let $(y_{u_1}^j, y_{u_2}^j, \ldots, y_{u_k}^j) \in [q]^k$ be a random sample from the distribution $\mu_{z_{u_1}^j} \times \mu_{z_{u_2}^j} \times \cdots \times \mu_{z_{u_k}^j}$.
    
    (d) Let the corresponding hyperedge in $\mathcal{D}$ be
    
    $$\left(\left(B(u_1), (y_{u_1}^1, y_{u_1}^2, \ldots, y_{u_1}^r)\right), \left(B(u_2), (y_{u_2}^1, y_{u_2}^2, \ldots, y_{u_2}^r)\right), \ldots, \left(B(u_k), (y_{u_k}^1, y_{u_k}^2, \ldots, y_{u_k}^r)\right)\right)$$
    
    and the constraint corresponding to it be $A_e$.

**Properties of the Dictatorship Function.**

**Definition 5.9.** An assignment $f : [m] \times [q]^r \mapsto [q]$ is said to be a dictator if there exists an $j \in [r]$ such that $f(b, z) = z_j$.

Given $b \in [m]$, let $f_b$ denote the restriction of $f$ to $(b, [q]^r)$. Given $p \in \Delta_q$, let $z \leftarrow \mu_p^r$ denote a string in $[q]^r$ drawn from the product distribution $\mu_p^r$. We can think of an assignment $f_b : [q]^r \mapsto [q]$ also as a a function from $[q]^r$ to $\Delta_q$ (where the value $i \in [q]$ gets associated with $e_i \in \Delta_q$). Given $j \in [r], b \in [m]$, the degree $d$-influence of the $j$-th coordinate of $f_b$, $\text{Inf}_j^{\leq d}(f_b)$ is defined as in Section 6.4.1.

**Definition 5.10.** Given $\tau, d \geq 0$, an assignment $f : [m] \times [q]^r \mapsto \Delta_q$ is said to be $(\tau, d)$-pseudo-random if for every $b \in [m]$ and every $j \in [r]$, $\text{Inf}_j^{\leq d}(f_b) \leq \tau$.

Now we state the completeness and soundness properties of the dictatorship function.

**Lemma 5.11.** *(Completeness.)* Let $f : [m] \times [q]^r \mapsto [q]$ be a dictator. Then $f$ satisfies all the constraints of $\mathcal{D}$ and $\sum_{v \in V(\mathcal{D})} w_{\mathcal{D}}(v) \cdot f(v) \leq \text{val}(\mathcal{I}, x) + O_q(\delta)$.

**Theorem 5.12.** *(Soundness.) For every small enough $\delta > 0$, there exists a $d, \tau$ such that if $f : [m] \times [q]^r \mapsto [q]$ satisfies all the constraints of $\mathcal{D}$ and is $(\tau, d)$-pseudo-random, then*

$$\sum_{v \in V(\mathcal{D})} w_{\mathcal{D}}(v) \cdot f(v) \geq \text{round}_q(\mathcal{I}, x, \varepsilon) - \Omega_q(\delta).$$

We note that the proof of this theorem is almost identical to that of Theorem 4.10. Here we outline the changes to be made in the $q > 2$ case. The crux of the proof is to show that given a subset $f : [m] \times [q]^r \mapsto [q]$, satisfying all the constraints in $\mathcal{D}$, one can find a function $g : [m] \times [q]^r \mapsto [q]$ such that $g$ depends only on $b \in [m]$ ($g$ is constant on every cube), $g$ satisfies all the constraints in $\mathcal{D}$ and $\sum_{v \in \mathcal{D}} w_{\mathcal{D}}(v) \cdot g(v) \leq \sum_{v \in \mathcal{D}} w_{\mathcal{D}}(v) \cdot f(v) + O_q(\delta)$. To obtain $g$, the strategy would be to look at the cube $\{b\} \times [q]^r$, find the smallest $i \in [q]$ such that

$$\mathop{\mathbf{E}}_{z \leftarrow \mu_{\tilde{p}_b}} f_{b,i}(z) \geq \delta.$$

Here, $f_{b,i} : [q]^r \mapsto \{0, 1\}$ denotes the indicator of the event that $f_{b,i}(z) = 1$ if $f(b, z) = i$ and 0 otherwise. We would then define $g(b, z) = i$ for all $z \in [q]^r$. Thus, $g$ has the desired form. Using the fact that $f$ is $d, \tau$-pseudorandom, one can now appeal to the Invariance Principle in the $q$-ary setting, as in Theorem 4.10, to obtain that $g$ still satisfies all the constraints in $\mathcal{D}$. Note that $\sum_{v \in V(\mathcal{D})} w_{\mathcal{D}}(v) \cdot g(v) \leq \sum_{v \in V(\mathcal{D})} w_{\mathcal{D}}(v) \cdot f(v) + \delta \cdot q$. Further, since $g$ corresponds to integral assignments to every cube, this corresponds to one of the assignments that the rounding algorithm iterates through on the input $(\mathcal{I}, x, \varepsilon)$. This implies that $\mathsf{round}_q(\mathcal{I}, x, \varepsilon)$ is at most $\sum_{v \in V(\mathcal{D})} w_{\mathcal{D}}(v) \cdot g(v)$. This completes the sketch of proof of the theorem.

**Unique Games Conjecture based Hardness Results.** The following are equivalent to Theorem 1.1 and Corollary 1.2.

**Theorem 5.13.** *Let $\Pi$ be a $k$-sm-CSP on alphabet $[q]$, $\varepsilon > 0$ for $k = O(1)$, and $\mathcal{I}$ be an instance of $\Pi$. Then for every $\delta > 0$, assuming the UGC, it is NP-hard to distinguish instances of $\Pi$ with optimal less than $\mathsf{lp}(\mathcal{I}) + \varepsilon + O_q(\delta)$ from those with optimal more than $\mathsf{round}_q(\mathcal{I}, x^\star, \varepsilon) - \Omega_q(\delta)$. Here $x^\star$ is the optimal LP solution for $\mathcal{I}$.*

**Corollary 5.14.** *Let $\Pi$ be in the class $k$-sm-CSP on alphabet $[q]$ and $\mathcal{I}$ be an instance of $\Pi$. Then for every $\delta > 0$, assuming the UGC, it is NP-hard to distinguish instances of $\Pi$ with optimal less than $\mathsf{lp}(\mathcal{I}) + O_q(\delta)$ from those with optimal more than $\mathsf{opt}(\mathcal{I}) - \Omega_q(\delta)$.*

The proofs follow verbatim along the lines of their counterparts for the binary case by converting the dictatorship function describe above.

**Acknowledgments** The authors would like to thank Oded Regev for bringing the paper [AKS09] to our notice and also observing that every problem in the class $k$-sm-CSP over the alphabet $\{0, 1\}$ can be reduced to a Hypergraph Vertex Cover problem in the approximation preserving sense.

## 6  Appendix

### 6.1  Comparison of our LP with standard LP's for Vertex Cover and Hypergraph Vertex Cover

We compare the LP relaxation of Figure 2 to the standard LP's for the Vertex Cover and $k$-Hypergraph Vertex Cover problems. For the Vertex Cover problem, the two LP's are equivalent. Fix an instance $\mathcal{I}$ of the Vertex

Cover problem. For an edge $e = uv$, the set $A_e = \{(0, 1), (1, 0), (1, 1)\}$. Hence, an element of ConvexHull($A_e$) can be written as $\lambda_1(0, 1) + \lambda_2(1, 0) + \lambda_3(1, 1) = (\lambda_2 + \lambda_3, \lambda_1 + \lambda_3)$, where $\lambda_i \geq 0$ for $i = 1, 2, 3$ and $\sum_{i=1}^{3} \lambda_i = 1$. Now, $x_u = \lambda_2 + \lambda_3, x_v = \lambda_1 + \lambda_3$ implies that $x_u + x_v \geq 1$. Conversely, it is easy to check that given values $x_u, x_v \geq 0$ such that $x_u + x_v \geq 1$, one can find corresponding $\lambda_i$ values. Thus, the standard LP and our LP for Vertex Cover problem are equivalent.

We now show that the LP of Figure 2 is stronger than the usual LP relaxation for $k$-Hypergraph Vertex Cover. An instance of $k$-Hypergraph Vertex Cover is specified by a ground set $V$ and a set $E \subseteq V^k$. Each vertex $u \in V$ has weight $w_u$ and the goal is to find a minimum weight subset of $V$ which contains at least one element from each hyperedge in $E$. The usual LP for $k$-Hypergraph Vertex Cover is written as follows :

$$\text{minimize} \quad \sum_{v \in V} w_v x_v \qquad (7)$$

$$\text{subject to} \quad \forall_{e = u_1, u_2, \ldots, u_k \in E} \quad x_{u_1} + \cdots + x_{u_k} \geq 1 \qquad (8)$$

$$\forall_{v \in V} \quad x_v \geq 0 \qquad (9)$$

The set $A_e$ for this problem contains all the elements of $\{0, 1\}^k$ except the all zero vector $(0, \ldots, 0)$. An element $(x_{u_1}, \ldots, x_{u_k}) \in A_e$ has the property that $x_{u_1} + \cdots + x_{u_k} \geq 1$ and so the same holds for any element in ConvexHull($A_e$). Hence, constraints (2) imply constraints (8).

## 6.2 Details of Section 3

We prove Fact 3.2 which we restate below.

**Fact 6.1.** *Let $x$ be a feasible solution to LP($\mathcal{I}$). Then*

1. *$x^\varepsilon$ is feasible for LP($\mathcal{I}$).*

2. *val($\mathcal{I}, x^\varepsilon$) $\leq$ val($\mathcal{I}, x$) + $\varepsilon$.*

*Proof.* We first prove the first statement. It is enough to prove this for $x'$ where $x'$ differs from $x$ on only one coordinate $u$. Fix an edge $e = (u_1, \ldots, u_k)$ and without loss of generality assume that $u = u_1$. Let $\lambda_\sigma, \sigma \in A_e$ be the coefficients in the convex combination of vectors in $A_e$ which yield $(x_{u_1}, \ldots, x_{u_k})$. Let $A'_e$ be the set of $\sigma$ for which $\sigma_1 = 0$.

For each $\sigma \in A'_e$, define $m(\sigma)$ as vector which is same as $\sigma$ except that $\sigma'_1 = 1$. Clearly, $m(\sigma) \in A_e$ as well. Now consider the vector $\sum_{\sigma \notin A'_e} \lambda_\sigma \sigma + \sum_{\sigma \in A'_e} \lambda_\sigma m(\sigma)$. This is equal to $(1, x_{u_2}, \ldots, x_{u_k})$. Thus, we have shown that the vector $x''$ which is identical to $x$ except that $x''_u = 1$ is feasible to LP($\mathcal{I}$). Now note that $x'$ is a convex combination of $x$ and $x''$. Hence, the claim follows.

We now prove the second statement. Since $x^\varepsilon_u \leq x_u + \varepsilon$, we get that

$$\text{val}(\mathcal{I}, x^\varepsilon) = \sum_u w_u x^\varepsilon_u \leq \sum_u w_u x_u + \varepsilon \sum_u w_u = \text{val}(\mathcal{I}, x) + \varepsilon.$$

□

## 6.3 Details of Section 4.1

We prove Lemma 4.6 which is restated below.

**Lemma 6.2.** *Let $A$ be a upward monotone subset of $\{0, 1\}^k$ and let $(x_1, x_2, \ldots, x_k) \in$ ConvexHull($A$) such that each $x_u$ is an integral multiple of $\varepsilon$ (assume that $1/\varepsilon$ is an integer). Then, there is a distribution $P$ over $A$ such that the minimum probability of any atom in $P$ is at least $\frac{\varepsilon}{2^k!}$.*

*Proof.* We define a variable $\lambda_\sigma$ for every $\sigma \in A_e$. We want to find a solution to the following :

$$\forall_{j\in\{1,\ldots,k\}} \quad \sum_{\sigma \in A_e} \lambda_\sigma \sigma_{u_j} = x_{u_j} \tag{10}$$

$$\sum_{\sigma \in A_e} \lambda_\sigma = 1 \tag{11}$$

$$\forall_{\sigma \in A_e} \quad \lambda_\sigma \geq 0 \tag{12}$$

We know that there is a feasible solution to this set of constraints. A vertex solution corresponds to the unique solution obtained by subset of the constraints (where inequality is replaced by equality). For a such a system of equations, $\mathbf{A} \cdot \lambda = \mathbf{b}$, we observe that all entries in $\mathbf{A}$ are 0-1 and all entries in $\mathbf{b}$ are integral multiples of $\varepsilon$. The determinant of $\mathbf{A}$ is at most $2^k!$. The value of $\lambda_\sigma$ (by Cramer's rule) is the ratio of determinants of two matrices – the matrix $\mathbf{A}$ with one of the columns replaced by $\mathbf{b}$, and the matrix $\mathbf{A}$. Since each entry of $\mathbf{b}$ is a multiple of $\varepsilon$, it is easy to check that the determinant of the former matrix is either 0 or at least $\varepsilon$. This implies the lemma. □

## 6.4 Details of Dictatorship Function (Section 4.2)

We give a brief introduction to the Invariance Principle in this section. Then we complete the soundness analysis (Theorem 4.10) of the dictatorship function by giving proof of Fact 4.11.

### 6.4.1 Gaussian Spaces and Mossel's Invariance Principle

**Measure Spaces.** We will be concerned with real valued functions in two measures spaces.

1. **$p$-biased measure space.** For $p \in [0, 1]$, the $p$-biased measure on $\{0, 1\}^r$ is denoted by $\mu_p$ where for $x = (x_1, \ldots, x_r) \in \{0, 1\}^r$, $\mu_p(x) \stackrel{\text{def}}{=} p^{|\{i:x_i=1\}|}(1-p)^{|\{i:x_i=0\}|}$. For $f, g : \{0, 1\} \mapsto \mathbb{R}$, define the following inner product: $\langle f, g \rangle_p := \mathbf{E}_{x \leftarrow \mu_p}[f(x)g(x)]$.

2. **Gaussian measure space.** We will denote by $\gamma$ as the Gaussian measure on $\mathbb{R}^r$ with density $\gamma(x) \stackrel{\text{def}}{=} (2\pi)^{-r/2} e^{-\|x\|^2/2}$ for $x \in \mathbb{R}^r$. For a function $f : \mathbb{R}^r \mapsto \mathbb{R}$ we will denote by $\mathbf{E}_\gamma[f] \stackrel{\text{def}}{=} \int_{\mathbb{R}^r} f(x)\gamma(x)dx$. We will restrict ourselves to $f \in L^2(\mathbb{R}^r, \gamma)$, i.e., $f$ such that $\mathbf{E}_\gamma[f^2] < \infty$. For $f, g : \mathbb{R}^r \mapsto \mathbb{R}$, define the following inner product: $\langle f, g \rangle_\gamma := \mathbf{E}_{x \leftarrow \gamma}[f(x)g(x)]$.

**Gaussian stability.** For $\rho \in [-1, 1]$, we denote by $U_\rho$ the Ornstein-Uhlenbeck operator $U_\rho$ which acts on $L^2(\mathbb{R}^r, \gamma)$ as $U_\rho f(x) \stackrel{\text{def}}{=} \mathbf{E}_{y \leftarrow \gamma}[f(\rho x + \sqrt{1-\rho^2} y)]$. It is easy to see that $\mathbf{E}_\gamma[U_\rho f] = \mathbf{E}_\gamma[f]$. For $0 \leq \mu \leq 1$, let $F_\mu : \mathbb{R} \mapsto \{0, 1\}$ denote the function $F_\mu(x) = 1_{\{x<t\}}$, where $t$ is chosen is a way such that $\mathbf{E}_\gamma[F_\mu] = \mu$.

**Definition 6.3.** Given $\mu, \nu \in [0, 1]$ and $\rho \in [-1, 1]$ define

$$\Gamma_\rho(\mu, \nu) \stackrel{\text{def}}{=} \langle F_\mu, U_\rho(1 - F_{1-\nu}) \rangle_\gamma.$$

For a vector $(\rho_1, \rho_2, \ldots, \rho_{k-1}) \in [-1, 1]^{k-1}$ and $\mu_1, \mu_2, \ldots, \mu_k$ we recursively define

$$\Gamma_{(\rho_1,\rho_2,\ldots,\rho_{k-1})}(\mu_1, \mu_2, \ldots, \mu_k) \stackrel{\text{def}}{=} \Gamma_{\rho_1}(\mu_1, \Gamma_{(\rho_2,\ldots,\rho_{k-1})}(\mu_2, \ldots, \mu_k)).$$

When the $\rho_i$ are all equal to $\rho$, and the $\mu_i$ are all equal to $\mu$, we will use $\Gamma_\rho^k(\mu)$ to denote the term on left hand side above.

We will also need the following simple facts.

**Fact 6.4.** *For every $\theta \in (0, 1)$ and $\rho = 1 - \lambda \in (0, 1)$, $\Gamma_\rho(\theta, \theta) \geq \theta^{1/\lambda}$.*

We use the above fact iteratively to obtain the following bound.

**Fact 6.5.** *For every $\theta \in (0, 1)$ and $\rho = 1 - \lambda \in (0, 1)$, $\Gamma_\rho^k(\theta) \geq \theta^{1/\lambda^k}$.*

**Product Spaces and Influences.** Let $(\Omega^1, \mu_1), \ldots, (\Omega^r, \mu_r)$ be probability spaces and let $(\Omega, \mu)$ denote the product space $(\prod_{i=1}^r \Omega^i, \prod_{i=1}^r \mu_i)$. Let $f = (f_1, \ldots, f_k) : \Omega \mapsto \mathbb{R}^k$. The influence of the $i$-th coordinate on $f$ is defined as

$$\mathbf{Inf}_i(f) \stackrel{\text{def}}{=} \sum_{1 \leq j \leq k} \mathop{\mathbf{E}}_{x=(x_1,\ldots,x_r) \leftarrow \mu} [\mathbf{Var}_{\mu_i}[f_j(x)|x_1, \ldots, x_{i-1}, x_{i+1}, \ldots, x_r]].$$

Here, fixing $(x_1, \ldots, x_{i-1})$ and $(x_{j+1}, \ldots, x_r)$, $f_j(x)$ is just a function of $x_i$, call it $g$. Hence, the variance $\mathbf{Var}_{\mu_i}[g] \stackrel{\text{def}}{=} \mathbf{E}_{\mu_i}[g^2] - (\mathbf{E}_{\mu_i}[g])^2$. When $\Omega^1 = \cdots = \Omega^r$ and $f$ is a boolean function, let $\{\hat{f}_S : S \subseteq \{0, 1\}^r\}$ denote the Fourier coefficients of $f$ with respect to the product measure $\mu$. Define the degree-$d$ influence of the $i$-th coordinate on $f$, $\mathbf{Inf}_i^{\leq d}(f)$, as $\sum_{S: i \in S, |S| \leq d} \hat{f}_S^2$.

**Correlated Spaces.** We now consider a correlated space $P$ on $\Omega^1 \times \Omega^2 \times \cdots \times \Omega^k$ with a probability measure $\mu$. A function $f : \Omega^1 \times \Omega^2 \times \cdots \times \Omega^k \mapsto \mathbb{R}$ is in $L^2(P)$ if $\mathbf{E}_{x \leftarrow \mu}[f^2(x)] < \infty$. We consider $L^2(P)$ as a vector space over the reals of all functions in $L^2(P)$ where addition of two functions is defined as point-wise addition. We denote by $\mathbf{Var}_\mu[f] \stackrel{\text{def}}{=} \mathbf{E}_\mu[f^2] - (\mathbf{E}_\mu[f])^2$, and $\mathbf{Cov}[f, g] \stackrel{\text{def}}{=} \mathbf{E}_\mu[fg] - \mathbf{E}_\mu[f]\mathbf{E}_\mu[g]$.

**Definition 6.6.** For any two linear subspaces $A$ and $B$ of $L^2(P)$, we define the correlation between $A$ and $B$ by

$$\rho(A, B; P) \stackrel{\text{def}}{=} \sup\{\mathbf{Cov}[f, g] : f \in A, g \in B, \mathbf{Var}[f] = \mathbf{Var}[g] = 1\}.$$

Now we can define the correlation, $\rho(P)$ of a correlated space $P$ over $\Omega^1 \times \Omega^2 \times \cdots \times \Omega^k$.

**Definition 6.7** (Correlation)**.** *The correlation $\rho(P)$ of a space $P$ over $\Omega^1 \times \Omega^2 \times \cdots \times \Omega^k$ is defined as*

$$\rho(P) \stackrel{\text{def}}{=} \max_{j=1}^k \left\{\rho(\Omega^1 \times \cdots \times \Omega^{j-1} \times \Omega^{j+1} \times \cdots \times \Omega^k, \Omega^j; P)\right\}.$$

We will use the following theorems in [Mos08].

**Theorem 6.8** (Cheeger's Inequality)**.** *[Mos08] Let $(\Omega^1 \times \Omega^2, P)$ be two correlated spaces such that the probability of the smallest atom in $\Omega^1 \times \Omega^2$ is at least $\alpha > 0$. Define a bipartite graph $G = (\Omega^1, \Omega^2, E)$ where $(a, b) \in \Omega^1 \times \Omega^2$ satisfies $(a, b) \in E$ if $P(a, b) > 0$. Then if $G$ is connected then*

$$\rho(\Omega^1, \Omega^2; P) \leq 1 - \alpha^2/2.$$

**Theorem 6.9** (Invariance Principle)**.** *[Mos08] Let $(\prod_{j=1}^k \Omega_i^j, P_i), 1 \leq i \leq r$ be a sequence of correlated probability spaces such that for all $1 \leq i \leq r$ the minimum probability of any atom in $\prod_{j=1}^k \Omega_i^j$ is at least $\alpha$. Assume furthermore that there exists $\rho \in [0, 1]^{k-1}$ and $0 \leq \rho_0 < 1$ such that $\rho(\Omega_i^1, \ldots, \Omega_i^k; P_i) \leq \rho_0$ and $\rho(\Omega_i^{\{1,\ldots,j\}}, \Omega_i^{\{j+1,\ldots,k\}}; P_i) \leq \rho_j$ for all $i, j$. Then for all $\eta > 0$ there exists $\tau > 0$ such that if $f_j : \prod_{i=1}^r \Omega_i^j \to [0, 1]$. for $1 \leq j \leq k$ satisfy $\max_{i,j}(\mathbf{Inf}_i(f_j)) \leq \tau$ then*

$$\Gamma_\rho(\mathbf{E}[f_1], \ldots, \mathbf{E}[f_k]) - \eta \leq \mathbf{E}\left[\prod_{j=1}^k f_j\right]. \tag{13}$$

*One may take $\tau = \eta^{O\left(\frac{\log(1/\eta)\log(1/\alpha)}{(1-\rho)\eta}\right)}$.*

### 6.4.2 Completing the soundness proof of dictatorship function

We now give the proof of Fact 4.11 which will complete the proof of Theorem 4.10. We first state the statement of this Fact below.

**Fact 6.10.**
$$\mathbf{E}\left[\Pi_{u \in R_e} f^S_{B(u)}(z^{(u)})\right] > 0.$$

*Proof.* Let $\alpha \stackrel{\text{def}}{=} \frac{\delta^k \varepsilon}{2^k!}$ and $Q_e \stackrel{\text{def}}{=} M_\delta(P^x_e)$. We know that the minimum probability of any atom in $Q_e$ is at least $\alpha$. Using Lemma 6.11 proved below, we have that the correlated space induced by $Q_e$ on $R_e$ has correlation at most $1 - \alpha^2$. Finally, using Fact 6.5, we know that the quantity $\Gamma_{(\rho,\ldots,\rho)}(\delta,\ldots,\delta)$ is at least $\delta^{1/\alpha^k} \stackrel{\text{def}}{=} \beta$. Thus, using Theorem 6.9 with $\eta = \beta/2$, we obtain $\tau = \tau(\eta)$ and $d = d(\eta)$ such that if for every $i$, $\text{Inf}^{\leq d}_i(f^S_b) \leq \tau$, then

$$\mathbf{E}\left[\Pi_{u \in R_e} f^S_{B(u)}\right] \geq \beta - \beta/2 \geq \beta/2 > 0.$$

□

It remains to prove the following lemma.

**Lemma 6.11.** *Consider an edge $e = (u_1, u_2, \ldots, u_k) \in E$. Let $\Omega^i$ denote the set $\{0, 1\}$, $i = 1, \ldots, k$. Let $S_1, S_2$ be two non-empty disjoint subsets of $\{u_1, u_2, \ldots, u_k\}$. Then the correlated space induced by $Q_e$ on $(\times_{i \in S_1} \Omega^i) \times (\times_{i' \in S_2} \Omega^{i'})$ has correlation at most $1 - \left(\frac{\delta^k \varepsilon}{2^k!}\right)^2$.*

*Proof.* Let $U$ denote the set $S_1 \cup S_2$. Let $Q^{S_1}_e, Q^{S_2}_e, Q^U_e$ be the distributions obtained by restricting $Q_e$ to $S_1, S_2$ and $U$ respectively. Let $\Omega^{S_1}$ denote the elements in $\times_{j \in S_1} \Omega^j$ with non-zero probabilites associated by $Q^{S_1}_e$. Define $\Omega^{S_2}$ similarly. Construct a bipartite graph on $\Omega^{S_1} \times \Omega^{S_2}$ where we have an edge $(a, b)$ if $Q^U_e((a, b)) > 0$. We now argue that this graph is connected. Indeed, let $a \in \Omega^{S_1}$ and $b \in \Omega^{S_2}$. Since $Q^{S_1}_e(a) > 0$, $(a, \mathbf{1})$ is an edge in this graph, where $\mathbf{1}$ is the all 1 vector of the appropriate dimension. Similarly, $(\mathbf{1}, b)$ and $(\mathbf{1}, \mathbf{1})$ are edges in this graph. It follows that the vertices $a$ and $b$ are connected. Note that for every edge $(a, b)$ in this graph, $Q^U_e((a, b)) \geq \frac{\delta^k \varepsilon}{2^k!}$. We now invoke Lemma 6.8 to finish the proof. □

## 6.5 The Reduction

In this section, we give the reduction from UNIQUE GAMES to a problem $\Pi$ in the class $k$-sm-CSP. We first state the version of UGC on which our results rely.

**Definition 6.12** (UNIQUE GAMES). *An instance $\mathcal{U} = (G(U, A), [r], \{\pi_e\}_{e \in A}, \text{wt})$ of UNIQUE GAMES is defined as follows: $G = (U, A)$ is a bipartite graph with set of vertices $U = U_{\text{left}} \cup U_{\text{right}}$ and a set of edges $A$. For every $e = (v, w) \in E$ with $v \in U_{\text{left}}, w \in U_{\text{right}}$, there is a bijection $\pi_e : [r] \mapsto [r]$, and a weight $\text{wt}(e) \in \mathbb{R}_{\geq 0}$. We assume that $\sum_{e \in E} \text{wt}(e) = 1$. The goal is to assign one label to every vertex of the graph from the set $[r]$ which maximizes the weight of the edges satisfied. A labeling $\Lambda : U \mapsto [r]$ satisfies an edge $e = (v, w)$, if $\Lambda(w) = \pi_e(\Lambda(v))$.*

The following notations will be used in the hardness reduction and we state them here.

**Notations.**

1. For a vertex $v \in U$, $\Gamma(v)$ is the set of edges incident to $v$.

2. For a vertex $v \in U$, define $p_v \stackrel{\text{def}}{=} \sum_{e \in \Gamma(v)} \text{wt}(e)$. This gives a probability distribution over the vertices in $U_{\text{left}}$ (or $U_{\text{right}}$).

We now state the Strong UGC which was shown by Khot and Regev [KR08] to be equivalent to the UGC [Kho02].

**Conjecture 6.13** (Strong UGC). *For every pair of constants $\eta, \zeta > 0$, there exists a sufficiently large constant $r := r(\eta, \zeta)$, such that it is NP-hard to distinguish between the following cases for an instance $\mathcal{U} = (G(U, A), [r], \{\pi_e\}_{e \in A}, \text{wt})$ of* Unique Games*:*

- **YES**: *There is a labeling $\Lambda$ and a set $U_0 \subseteq U_{\text{left}}$ of vertices, $\sum_{u \in U_0} p_u \geq (1 - \eta)$, such that $\Lambda$ satisfies all edges incident to $U_0$.*

- **NO**: *There is no labeling which satisfies a set of edges of total weight value more than $\zeta$.*

Now we describe the reduction from Unique Games instance to our problem. The reduction shall use the instance $\mathcal{D}$ of $\Pi$ produced by $\text{Dict}^{\Pi}_{\mathcal{I},x,m}(r, \delta)$. Here $x$ is the $\varepsilon$-perturbed solution corresponding to an optimal solution $x^\star$ to $\text{lp}(\mathcal{I})$.

**Input Instance :** The input to the reduction is an instance $\mathcal{U} = (G(U, A), [r], \{\pi_e\}_{e \in A}, \text{wt})$ of Unique Games problem as defined in Definition 6.12. Recall that $G$ is a bipartite graph with $U = U_{\text{left}} \cup U_{\text{right}}$, and the edge weights wt induce probability distribution $p_v$ over vertices in $U_{\text{left}}$.

**Output Instance :** The output instance $\mathcal{F}$ of $\Pi$ is as follows :

1. Vertex Set $V(\mathcal{F}) = U_{\text{left}} \times V(\mathcal{D})$, i.e., we place a copy of $V(\mathcal{D})$ at each vertex of $U_{\text{left}}$. We shall index a vertex by $(u, b, y)$ where $u \in U_{\text{left}}$ and $(b, y) \in V(\mathcal{D})$.

2. Vertex Weights The weight of a vertex $(u, b, y)$ is
$$w_{\mathcal{F}}((u, b, y)) = p_u \cdot w_{\mathcal{D}}((b, y)).$$

3. Hyper-edges For every hyperedge $e = \left((b^1, y^1), (b^2, y^2), \ldots, (b^k, y^k)\right)$ in $\mathcal{D}$, we add the following edges to $\mathcal{F}$
   – for each vertex $u \in U_{\text{right}}$ and all sets of $k$ neighbors, $u^1, \ldots, u^k$ (with repetition) of $u$, we add the hyperedge $\left((u^1, b^1, y^1 \circ \pi^u_{(u,u^1)}), \ldots, (u^k, b^k, y^k \circ \pi^u_{(u,u^k)})\right)$ to $\mathcal{F}$. The constraint for the these edges is the same as that for $e$.

**Completeness.**

**Theorem 6.14.** *Suppose there is a labeling $\lambda$ for $\mathcal{U}$ and a subset $U_0$ of $U_{\text{left}}$, $\sum_{v \in U_0} p_v \geq 1 - \eta$, such that $\lambda$ satisfies all edges incident on $U_0$. Then there is a subset of vertices in $\mathcal{F}$ which satisfy all the constraints in $\mathcal{F}$ and has weight at most $\text{lp}(\mathcal{I}) + \delta + \varepsilon + \eta$.*

*Proof.* Consider the labeling $\lambda$. We now show how to pick a set $F$ of vertices from $V(\mathcal{F})$ which satisfies all the hyperedge constraints. For each $u \in U_0$, define $J_u$ as $\{(u, b, y) \in V(\mathcal{F}) : y_{\lambda_u} = 1\}$. For each $u \in U_{\text{left}} - U_0$, define $J'_u$ as the set $\{(u', b', y') \in V(\mathcal{F}) : u' = u\}$. Now define $F = \cup_{u \in U_0} J_u \bigcup \cup_{u \in U_{\text{left}} - U_0} J'_u$.

We now show that $F$ satisfies all hyperedge constraints. Fix a hyperedge $e = \left((b^1, y^1), \ldots, (b^k, y^k)\right)$ in $\mathcal{D}$. Let $u \in U_{\text{right}}$ and $u^1, \ldots, u^k$ be $k$ neighbors of $u$. Consider a corresponding edge $f = ((u^1, b^1, y^1 \circ \pi^u_{(u,u^1)}), \ldots, (u^k, b^k, y^k \circ \pi^u_{(u,u^k)}))$ in $\mathcal{F}$. Lemma 4.9 shows that the set $C_i = \{(b, z) : z_i = 1\}$ satisfies the edge constraint for $e$ for any $i$. Let us pick $i = \lambda_u$. It will be enough to prove that if $(b^l, y^l)$ satisfies $y^l_i = 1$, then the vertex $w = (u^l, b^l, y^l \circ \pi^u_{(u,u^l)})$ is in $F$. But this is indeed the case because if $u^l \in U_0$, then $\lambda_u = \pi^u_{(u,u^l)}(\lambda_{u^l})$. Therefore, $y^l \circ \pi^u_{(u,u^l)}$ has coordinate $\lambda_{u^l}$ equal to 1. Hence, $w \in J^l_u$. If $u^l \in U_{\text{left}} - U_0$, then we add $w \in J'_{u^l}$ trivially. Thus, we have shown that $F$ satsifies the edge constraint for the hyperedge $f$.

Let us now compute the weight of $F$. If $u \in U_0$, then Lemma 4.9 and Fact 3.2 show that the weight of $J_u$ is at most $p_u \cdot (\text{lp}(\mathcal{I}) + \varepsilon + \delta)$. If $u \notin U_0$, then the weight of $J'_u$ is $p_u$. Thus, the weight of $F$ is at most

$$(\text{lp}(\mathcal{I}) + \delta + \varepsilon) \sum_{u \in U_0} p_u + \sum_{u \notin U_0} p_u \leq \text{lp}(\mathcal{I}) + \delta + \varepsilon + \eta.$$

$\square$

**Soundness.**

**Theorem 6.15.** *Suppose there is a subset of vertices F which satisfies all the constraints in $\mathcal{F}$ and $w_{\mathcal{F}}(F) <$ round$(\mathcal{I}, x^\star, \varepsilon) - 2\delta$. Then there is a constant $\zeta(\varepsilon, \delta, k)$ such that there is a labeling for $\mathcal{U}$ for which the set of satisfied edges has weight at least $\zeta(\varepsilon, \delta, k)$.*

*Proof.* Consider a set $F$ satisfying the conditions of the theorem. Let $\mathbf{I}_F(\cdot)$ be the indicator function for $F$. For a vertex $u \in U_{\text{right}}$, let $N(u) \subseteq U_{\text{left}}$ denote the neighbors of $u$. Recall that every vertex of $\mathcal{F}$ can be written as $(w, z)$, where $w \in U_{\text{left}}$ and $z \in V(\mathcal{D})$. Since the distribution $\{p_w\}_{w \in U_{\text{left}}}$ is same as first picking a vertex $u \in U_{\text{right}}$ with probability $p_u$ and then picking a random neighbor of $u$ (according to edge weights), we get

$$w_{\mathcal{F}}(F) = \mathbf{E}_{u \in U_{\text{right}}} \mathbf{E}_{w \in N(u)} \mathbf{E}_{z \in V(\mathcal{D})} \mathbf{I}_F((w, z \circ \pi^u_{(u,w)})),$$

where $z$ is picked according to vertex weights in $\mathcal{D}$. For a vertex $u \in U_{\text{right}}$, let $G(u)$ denote the quantity

$$\mathbf{E}_{w \in N(u)} \mathbf{E}_{z \in V(\mathcal{D})} \mathbf{I}_F((w, z \circ \pi^u_{(u,w)})).$$

We can therefore state the condition of the Theorem as $\mathbf{E}_{u \in U_{\text{right}}} G(u) <$ round$(\mathcal{I}, x^\star, \varepsilon) - 2\delta$. Call a vertex $u \in U_{\text{right}}$ *good* if $G(u) <$ round$(\mathcal{I}, x^\star, \varepsilon) - \delta$. A simple averaging argument shows that the weight of good vertices is at least $\delta/2$.

Fix a good vertex $u$. Let $\mathcal{D}^{(u)}$ be a *copy* of the instance $\mathcal{D}$. We construct a solution $S^{(u)}$ for $\mathcal{D}^{(u)}$ as follows : for each $(b, y) \in V(\mathcal{D}^{(u)})$, we pick a random neighbor $u^i$ of $u$ according to edge weights wt in the instance $\mathcal{U}$. If $(u^i, b, y \circ \pi^u_{(u,u^i)}) \in F$, we add $(b, y)$ to $S^{(u)}$.

**Claim 6.16.** *$S^{(u)}$ satisfies all the constraints in $\mathcal{D}^{(u)}$.*

*Proof.* Let $e = \big((b^1, y^1), \ldots, (b^k, y^k)\big)$ be a hyperedge in $\mathcal{D}^{(u)}$. Suppose while constructing the set $S^{(u)}$, we decide to add $(b^i, y^i)$ to this set based on whether $(u^i, b^i, y^i \circ \pi^u_{(u,u^i)}) \in F$. Now observe that the instance $\mathcal{F}$ has the hyperedge $\big((u^1, b^1, y^1 \circ \pi^u_{(u,u^1)}), \ldots, (u^k, b^k, y^k \circ \pi^u_{(u,u^k)})\big)$. Since this hyperedge is satisfied by $F$, the claim follows. □

Note that $\mathbf{E}[S^{(u)}]$ is exactly $G(u)$, where the expectation is over the choice of random neighbors of $u$. For each vertex $w \in U_{\text{left}}$ and $b \in [m]$, define a 0-1 function $f_b^{F,w}$ on $\{0, 1\}^r$ as follows –

$$f_b^{F,w}(y) \stackrel{\text{def}}{=} \begin{cases} 1 & \text{if } (w, b, y) \notin F \\ 0 & \text{otherwise} \end{cases}$$

Note that $f_b^{F,w}$ is the indicator function for complement of $F$ for the set of vertices $\{(w, b, y) : y \in \{0, 1\}^r\}$. For the vertex $u$, we now define the function $f_b^{F,u}(y)$ which is the average of the corresponding functions for the neighbours of $u$.

$$f_b^{F,u}(y) \stackrel{\text{def}}{=} \mathbf{E}_{w \in N(u)} f_b^{F,w}\left(y \circ \pi^u_{(u,w)}\right).$$

Observe that $f_b^{F,u}(y) = \mathbf{P}[(u, b, y) \notin S^{(u)}]$, where the probability is over the choice of $S(u)$. Rest of the proof is very similar to the proof of Theorem 4.10 – the goal would be to prove the following statement :

**Lemma 6.17.** *There exist values $b \in [m], i \in [r]$ and constants $d, \tau$ depending on $\delta$ and $k$ only such that $\text{Inf}_i^{\leq d}(f_b^{F,u}) \geq \tau$.*

*Proof.* Let $\tau$ and $d$ be as in the proof of Fact 4.11. Suppose, for the sake of contradiction, that the statement of the lemma does not hold for these values of $\tau$ and $d$. Define

$$J^{(u)} \stackrel{\text{def}}{=} \left\{ b \mid \mathop{\mathbf{E}}_{z \leftarrow \mu_{\tilde{p}_b}} f_b^{F,u}(z) \leq \delta \right\}.$$

Let $T_J^{(u)} \stackrel{\text{def}}{=} \cup_{b \in J^{(u)}} (b, \{0,1\}^r)$. Fix a particular hyperedge $e = (a_1, a_2, \ldots, a_k) \in E$ of the instance $\mathcal{I}$ (which gets used in $\textsc{Dict}^\Pi_{\mathcal{I},x,m}(r,\delta)$). Let

$$R_e^{(u)} \stackrel{\text{def}}{=} \{a_i \in e \mid B(a_i) \notin J^{(u)}\}.$$

The following fact can be deduced in the same manner as the proof of Theorem 4.11.

**Fact 6.18.**
$$\mathbf{E} \, \Pi_{a \in R_e^{(u)}} f_{B(a)}^{F,u}(z^{(a)}) > 0,$$

*where the expectation is over the choice of $z^{(a)}$ from $\tilde{p}_{B(a)}$-biased measure on $\{0,1\}^r$.*

Fact 6.18 implies that there exist neighbors $u^a \in N(u)$ for each $a \in R_e^{(u)}$ such that

$$\mathbf{E} \, \Pi_{a \in R_e^{(u)}} f_{B(a)}^{F,u^a}(z^{(a)} \circ \pi^u_{(u,u^a)}) > 0.$$

Therefore, there exist values $\{z^{(a)} : a \in R_e^{(u)}\}$, such that $f_{B(a)}^{F,u^a}(z^{(a)} \circ \pi^u_{(u,u^a)}) = 1$ for all $a \in R_e^{(u)}$. But then consider the following choice of $S^{(u)}$: while considering $(B(a), z^{(a)})$, we pick the neighbor $u^a$ of $u$. Therefore, the set $S^{(u)}$ does not contain any element from the set $\{(B(a), z^{(a)}) : a \in R_e^{(u)}\}$. The set $S^{(u)}$ still satisfies all the constraints in $\mathcal{D}^{(u)}$ (Lemma 6.16). Hence, there exists a hyperedge $\left((b^1, z^1), \ldots, (b^k, z^k)\right)$ in $\mathcal{D}^u$ corresponding to $e$ which gets satisfied by $T_J^{(u)}$ only. Thus, all hyperedges in $\mathcal{D}^{(u)}$ are satisfied by $T_J^{(u)}$. But $w_\mathcal{D}(T_j^{(u)}) \geq \textsf{round}(\mathcal{I}, x^\star, \varepsilon)$. Further, we know that

$$\mathbf{E} \, w_\mathcal{D}(S^{(u)}) = G(u) \geq w_\mathcal{D}(T_J^{(u)}) - \delta.$$

This contradicts the fact that $\mathbf{E} \, w_{\mathcal{D}^{(u)}} S^{(u)} < \textsf{round}(\mathcal{I}, x^\star, \varepsilon) - \delta$. Hence, the lemma is true. □

The rest of the argument to complete the theorem follows standard arguments, see e.g. [Rag08]. □

**Choice of Parameters and Proof of Theorem 1.1** Given parameters $\varepsilon$ and $\delta$, and a constant value $k$, we first pick $\eta$ according to the proof of Fact 4.11 described in Section 6.4.2. The parameter $\eta$ can be assumed to be much less than $\delta$ (otherwise we can just set $\eta$ to be $\delta$). This yields the parameters $\tau$ and $d$ as specified by Lemma 6.17. Then, as in [Rag08], $\zeta(\varepsilon, \delta, k)$ mentioned in the statement of Theorem 6.15 depends on $\tau, d, \delta$, and hence, on $\varepsilon, \delta$ and $k$ only. Now we pick the label size $r$ of the Unique Games instance $\mathcal{U}$ to be large enough such that we get a gap of $1 - (\eta + \varepsilon + \delta)$ versus $\zeta$ in the UGC. Theorems 6.14 and 6.15 now imply Theorem 1.1 if we pick the values $\delta$ and $\varepsilon$ in the reduction to be half of the ones mentioned in the theorem.